\documentclass[aps,prx,twocolumn,floatfix,amssymb]{revtex4-2}

\usepackage{graphicx,dcolumn,bm,xstring,braket,amsmath,comment,nicefrac,placeins,amssymb}
\usepackage[unicode=true,pdfusetitle,bookmarks=true,bookmarksnumbered=true,bookmarksopen=false, breaklinks=false,pdfborder={0 0 0},pdfborderstyle={},backref=false,colorlinks=false, pdftitle={Atomic interferometer based on optical tweezers}]{hyperref}

\begin{document}

\title{Atomic clock interferometry using optical tweezers}

\author{Ilan Meltzer}
\author{Yoav Sagi}
\email[Electronic address: ]{yoavsagi@technion.ac.il}
\affiliation{Physics Department and Solid State Institute, Technion - Israel Institute of Technology, Haifa 32000, Israel}

\date{\today}

\begin{abstract}
Clock interferometry refers to the coherent splitting of a clock into two different paths and recombining in a way that reveals the proper time difference between them. Unlike the comparison of two separate clocks, this approach allows testing how non-flat spacetime influences quantum coherence.  Atomic clocks are currently the most accurate time keeping devices. Here we propose using optical tweezers to implement clock interferometry. Our proposed clock interferometer employs an alkaline-earth-like atom held in an optical trap at the magic wavelength. Through a combination of adiabatic, tweezer-based, splitting and recombining schemes and a modified Ramsey sequence on the clock states, we achieve a linear sensitivity to the gravitational time dilation. Moreover, the measurement of the time dilation is insensitive to relative fluctuations in the intensity of the tweezer beams. We analyze the tweezer clock interferometer and show that it is feasible with current technological capabilities. The proposed interferometer could test the effect of gravitational redshift on quantum coherence, and implement the quantum twin paradox. 
\end{abstract}

\maketitle

\section{Introduction}

Atom Interferometers (AIFs) are techniques for coherently manipulating spatial superpositions of atomic wavefunctions and measuring their phase difference. AIFs come in many forms, yet they all share the same basic steps and structure \cite{Cronin_2009}. The AIF consists of input and output ports, typically two of each type. The ports may be characterized by the outcome of different observables, such as position \cite{Nemirovsky_2023}, momentum \cite{Rasel_1995, Giltner_1995}, and angular momentum \cite{Raithel_2022}. A single atomic wavepacket populating one of the input ports is coherently split into two spatially separated wavepackets, so that each wavepacket is traveling on a different path. Each path may be under the influence of a different potential, so that each wavepacket acquires a different phase. After acquiring the phase, the wavepackets are coherently recombined. This process maps the phase difference between the wavepackets to population probabilities at the output ports. The populations are, in turn, measured to infer the phase difference. 

The most common type of AIFs is free space light-pulse interferometry, such as the Kasevich-Chu Interferometer (KCI) \cite{Kasevich_1991}. In such architectures, atoms in free fall are split and recombined in momentum space through the absorption of photons, thus creating two possible spatial trajectories. An alternative approach is to confine the wavepackets in a potential well during all or part of their trajectories. These types of AIFs, also known as guided AIFs, offer several potential advantages over traditional light-pulse AIFs, including arbitrary atom trajectories, precise positioning, long probing times, and compact experimental setups. An early demonstration of a guided AIF was conducted with a Bose-Einstein Condensate in a double-well potential \cite{Shin2004}. In recent years, several alternative architectures for guided AIFs have been proposed. Among the proposed methods for guided atomic interferometry, some techniques manipulate atomic trajectories using optical lattices \cite{Xu_2019,Raithel_2022}, while other utilize optical tweezers \cite{C_Ryu_2015,Nemirovsky_2023,GayathriniPremawardhana}. A significant challenge in guided interferometry lies in fulfilling the stringent requirement for the intensity stability of the confining optical potential, which is crucial to prevent differential phase noise.

AIFs, particularly KCI, have been successfully employed in numerous precise measurements \cite{Morel_2020,Hamilton_2015}. The fact that AIFs involve massive particles and can detect extremely small variations in force fields makes it an excellent choice for gravitational measurements. Indeed, AIFs have been used to test the weak equivalence principle \cite{Asenbaum_2020} and to determine the gravitational constant \cite{Fixler:2007is, Rosi_2014}. One central tenet of general relativity is the modification of proper time due to the metric. Proper time, denoted by $\tau$, is the time measured by an ideal clock moving with the reference frame. The theory of General Relativity predicts that clocks tick slower near a large mass, a prediction confirmed in several experiments comparing independent clocks positioned at different distances relative to Earth's mass \cite{Vessot_1980,Takamoto_2020,Chou_2010,Bothwell_2022,Herrmann_2018,Delva_2018}. 

In principle, time dilation can also be probed using AIF. As proper time is inherently tied to the metric in general relativity, observing its influence on the wavefunction's interference pattern could probe the regime in which both general relativity and quantum mechanics have a measurable effect \cite{Zych_2016}. As an example, some theories propose treating proper time as a quantum operator, with mass as its conjugate \cite{Greenberger_1970,Greenberger1970}. In these theories, the uncertainty in proper time is predicted to affect the visibility of the interference pattern \cite{Zych_2011}. Other proposals suggest interferometric measurements to test the connection between gravity and decoherence on macroscopic scales \cite{Pikovski_2015,Bassi_2017}.

The phase of a matter wave, $\phi$, is connected to it's proper time through \cite{Schleich_2013}:
\begin{equation} \label{eq1}
     \phi = \frac{mc^2}{\hbar}\tau \, ,
\end{equation}
where $m$ is the mass, $c$ is the speed of light, and $\hbar$ is the reduced Planck constant. This relation may lead one to conclude that it is possible to use the accumulated phase in AIF as a means to measure the difference of the proper time. This claim was made by Muller \textit{et al.} as they reinterpreted light-pulse interferometry experiments as a clock ticking in the Compton frequency, and as such, used its measured phase difference for calculating the gravitational red shift \cite{M_ller_2010}. However, it was claimed that the proper time cannot be measured by a particle with a single internal state \cite{Wolf_2011}. The phase difference in a light-pulse AIF in free fall is only due to the different phases of light pulses relative to the atoms accelerated by gravity, $g$ \cite{Schleich_2013}. To ascertain the proper time difference, it is necessary to employ a clock, which requires at least two internal states. While atomic clock interferometry (ACIF) has been previously realized with hyper-fine states \cite{Rosi_2017}, it has yet to be performed using states separated by optical transitions. Having a large transition frequency is a crucial requirement for a feasible measurement of the gravitational redshift.

A natural choice for clock interferometry is optical atomic clocks, which are the most accurate man-made timekeeping devices  \cite{Ludlow_2015}. These clocks are based on measuring the optical transitions between long-lived electronic states. Their operation relies on cold atom technology, namely the ability to precisely control with light the internal and external degrees of freedom of single atoms \cite{Metcalf_1999}. The atoms are held in an optical potential for long probing durations. To eliminate systematic shifts due to light shift, a `magic' wavelength is chosen for the optical potential where the two internal states have equal polarizability \cite{Ludlow_2015}. In most atomic clocks, the potential is generated by an optical lattice, but recently the use of optical tweezers has also been reported \cite{Young_2020}.

A step towards ACIF has been taken using the ${}^{1}S_0-{}^{3}P_0$ optical clock transition of ${}^{88}$Sr \cite{Hu_2017,Hu_2019,Rudolph_2020}. In these experiments, however, the splitting was accomplished using a single photon optical transition. Consequently, the atoms in each interferometer arm were in one of the clock states and not in a superposition of them. Therefore, they did not act as a clock in the sense required for measuring the proper time difference between the arms. Guided ACIFs hold a great promise for gravitational measurements, in particular for testing the Universality of the Gravitational Redshift (UGR) -- the principle that the gravitational time dilation is independent of the inner workings of the clock \cite{Roura_2020,Di_Pumpo_2021}. In most light pulse interferometers, the proper time difference between the arms is zero \cite{Schleich_2013,Giese_2019,Greenberger_2012}. Previous works were focused mainly on suggesting ACIFs based on light-pulse schemes that overcome this difficulty \cite{Di_Pumpo_2021,Roura_2020,Loriani_2019}. Guided interferometers, on the other hand, are inherently well-suited for measuring differences in proper time. \cite{Di_Pumpo_2021}. However, a concrete scheme for a guided ACIF using optical transitions has yet to be proposed.

In this work, we propose an ACIF scheme that uses optical pulses only to create a balanced superposition of the clock states in each arm, while the splitting process is achieved by using adiabatic tunneling transitions between optical tweezers. The scheme achieves linear sensitivity to gravitational time dilation, providing a significant signal in realistic experimental timescale. Importantly, we show that when the tweezer is at a magic wavelength, the interference signal is insensitive to relative intensity fluctuations between the tweezers. This property makes an interferometric measurement of the gravitational red shift with our proposed scheme feasible with current technological capabilities. The significance of this experiment lies in measuring a general gravity effect with a spatially separated coherent quantum state for the first time.

The structure of the paper is as follows. In section \ref{sec:Tweezer atomic interferometry} we briefly discuss tweezer atomic interferometry, previously proposed in Ref. \cite{Nemirovsky_2023} in relation to atoms with a single internal quantum state. In section \ref{sec:Tweezer clock interferometry} we generalize this scheme for tweezer clock interferometry, using unitary evolution calculations for the expected interference pattern. In section \ref{sec:Feasibility Analysis} we analyze the expected signal in a realistic experimental scenario and demonstrate that it can be measured. We conclude and give an outlook in section \ref{sec:Discussion}.

\section{Tweezer atomic interferometry}\label{sec:Tweezer atomic interferometry}

In Ref. \cite{Nemirovsky_2023}, we introduced a guided AIF scheme using optical tweezers. As the ACIF scheme we present in section \ref{sec:Tweezer clock interferometry} builds upon this concept, we briefly revisit it here for clarity. In the tweezer AIF, the atom's motion is controlled throughout the entire interferometric sequence by manipulating the position of the optical tweezers \cite{Florshaim2023}. The sequence starts with the atoms located in one of the tweezers. Coherent splitting and recombining are achieved by an adiabatic change of the traps' positions. We proposed two configurations for the splitters and combiners, utilizing either a two-tweezer or a three-tweezer setup.

In the two-tweezer scheme, the trap potential is written as
\begin{equation}
V(x,t) = -V_0 \left[e^{-2\frac{x-d\left(t\right)/2}{\sigma^2}}+\Big(1-\Delta\left(t\right)\Big) e^{-2\frac{x+d\left(t\right)/2}{\sigma^2}}\right] \, ,
\end{equation}
where $\sigma$ is the waist of the Gaussian beam creating each of the tweezers, $V_0$ is the potential depth, and $\Delta(t)$ is the relative depth difference (also referred to as detuning) between the two traps. Splitting of the wavefunction is achieved by changing the separation between the traps, $d(t)$, and the detuning in the following manner:
\begin{equation}
d(t) = \frac12 (d_{max}+d_{min}) + \frac12(d_{max}-d_{min})\cos \left(2\pi t/T\right)\\
\end{equation}
\begin{equation}
\Delta(t)=
\begin{cases}
        \Delta_{max}(1-2t/T) & \text{if }  t<T/2\\
        0 & \text{if } t \ge T/2
    \end{cases} \ \ ,
\end{equation}
where $T$ is the total splitting process time, and $d_{max}$ ($d_{min}$) is the initial (shortest) distance between the trap centers. The initial seperation is large enough such that tunneling is negligible. The initial detuning, $\Delta_{max}$, is chosen to be the largest possible value before eigenstates with different vibrational numbers cross. 

To explain how the splitting works, we consider a single atom which is initially placed in the deeper trap. The splitting sequence is slow enough to be adiabatic and consists of two parts. First, the traps are brought closer to a distance $d_{min}$ and simultaneously the detuning is decreased to zero. Then, the traps are moved back to their initial position while the detuning remains zero. The splitting occurs due to the adiabatic following of the atom's wavefunction -- at the beginning of the process, the atom is in the double-trap's ground state, located at the deeper trap. The adiabatic modification of the potential ensures that the final atomic wavefunction is still the trap potential's ground state, which is a balanced symmetrical split between the traps. If an atom is initially placed in the shallower trap, which corresponds to the first excited state of the initial Hamiltonian, it will end in the anti-symmetric split state.

After the splitting phase, the wavepacket in each of the interferometer arms acquires a phase at a different rate due to evolution under different potentials. At the end of the phase acquisition stage, the combiner, which is the time-reversal of the splitter, is applied. If the interferometer arms finish the process with a relative phase of 0 ($\pi$) radians, the combiner maps the atom position to the deep (shallow) trap. For any other phase, the final atomic wavefunction will be a superposition of populating each trap, thus achieving a phase-to-population mapping in the combiner step of an interferometer.

The three-tweezer splitting scheme introduced in Ref. \cite{Nemirovsky_2023} is similar to the two-tweezer scheme but also allows the detection of erroneous performance. Within the scope of this paper, the differences between two- and three-tweezer splitters are not important, as the operation in both scenarios is described by the same unitary matrix.

An important feature of the optical tweezer AIF scheme is that it can be executed with many atoms in a single run \cite{Nemirovsky_2023}. This is attributable to the fact that the splitting and recombining schemes can work successfully with different initial vibrational eigenstates as long as they are still adiabatic. Running an interferometric measurement with $N$ atoms in different vibrational states is equivalent to $N$ runs with one atom each, allowing for an improvement in the signal-to-noise ratio at a given number of repetitions. Preparation of each atom at a different eigenstate can be achieved by loading the tweezer with fermionic atoms in a single spin state \cite{Serwane_2011}.

The guided interferometer architecture utilizing optical tweezers presents several benefits compared to free-fall light-pulse interferometry. In light-pulse interferometers, the atoms are in free fall, necessitating a larger apparatus for longer interrogation times. Conversely, in guided interferometers, the interrogation time is not dependent on the apparatus size, enabling a more compact system. Another limitation of light-pulse interferometry is that the atoms are in free-fall. In the proposed tweezer AIF, the atom's position is entirely determined by the tweezer position, which allows for arbitrary trajectories during phase accumulation. Due to its enhanced sensitivity and arbitrary atom trajectory, the tweezer AIF can measure effects in classical and quantum gravity, e.g measuring the gravitational constant $G$ \cite{Tino_2021}, testing the Newtonian law at small distances \cite{Tino_2021}, measuring the gravitational redshift \cite{Di_Pumpo_2021} and searching for evidence of the gravitational-field quantization \cite{Marletto_2017,Bose_2017}. Most importantly, in the context of this paper, guided interferometers are sensitive to proper time differences between the arms \cite{Di_Pumpo_2021}, while existing light-pulse interferometers are inherently insensitive to proper time \cite{Greenberger_2012}. Moreover, an optical tweezer AIF working at a magic wavelength with an atomic clock is ideally suited to measure the gravitational time dilation, as we demonstrate in section \ref{sec:Tweezer clock interferometry}.

\section{Tweezer clock interferometry}\label{sec:Tweezer clock interferometry}

According to general relativity, the proper time, $\tau$, is determined by the metric, $g_{\mu\nu}$, according to
\begin{equation}
\tau = \int_\Gamma\,d\tau = \frac{1}{c^2}\int_\Gamma \sqrt{g_{\mu\nu}\,dx^\mu dx^\nu} \ \ ,
\end{equation}
where the integration is performed along a path $\Gamma$. Clocks that follow different paths may experience a difference in their proper time. In particular, if clocks are positioned at different locations under the influence of a gravitational field, they tick at a different rate. This prediction of general relativity was confirmed using two clocks at different heights \cite{Vessot_1980}. On the other hand, proper time is connected to matter-wave phase through Eq.(\ref{eq1}). This serves as a motivation for using a matter-wave interferometer as a probe in situations where both general relativity and quantum mechanics are relevant.

Optical tweezer AIFs, like those discussed in section \ref{sec:Tweezer atomic interferometry}, can be used to measure phases arising from proper time differences if the atom has at least two internal states. This is due to the fact that an atom with a single internal degree of freedom cannot produce a periodic signal necessary for measuring proper time \cite{Wolf_2011,Sinha_2011}. Interferometers employing only a single internal state can only measure effects stemming from a difference of the gravitational potential at different location, analogous to a gravitational Aharonov-Bohm effect \cite{Overstreet_2022}. Such observations were made using atomic and neutron interferometry \cite{Colella_1975}. Atomic clock interferometry was previously proposed as a method to measure proper time differences \cite{Roura_2020,Loriani_2019,Di_Pumpo_2021,Sinha_2011}. However, these proposals suffer from the same limitation of free falling AIF discussed in section \ref{sec:Tweezer atomic interferometry}. Previous discussions of trapped AIFs \cite{Roura_2020,Di_Pumpo_2021} were mostly concerned with calculating the redshift induced phase difference, without considering the questions of the splitting dynamics or the phase's measurement protocol. In what follows, we present a proposal for a guided ACIF, using optical tweezers at a magic wavelength, including an analysis of its available obeservables. Our scheme inherits the advantages of the tweezer AIF, namely long probing duration and high position accuracy. Additionally, the measurement of the proper time difference is insensitive to relative intensity fluctuations, which makes it applicable to current tweezer technology.

\begin{figure*}[!ht]
 \centering
	\includegraphics[width=\textwidth]{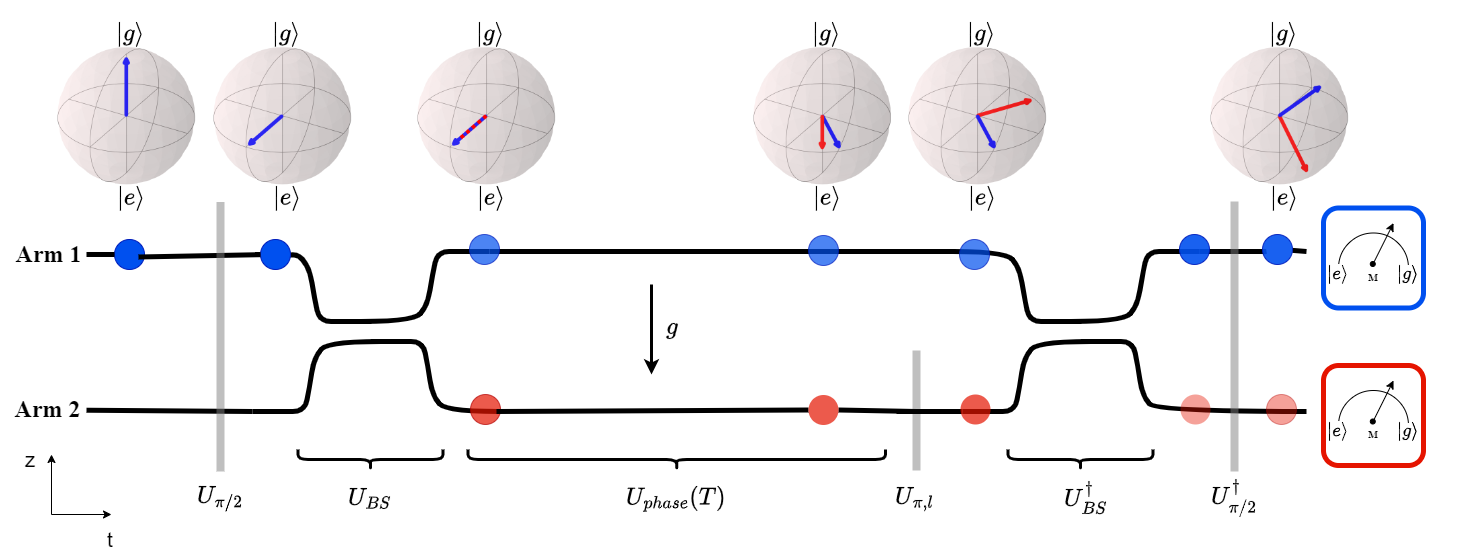}
	\caption{\textbf{The proposed scheme for interferometry with an atomic clock.} The black solid lines represent the tweezers' centers in 1D (vertical axis) as a function of time (horizontal axis). The upper trap is denoted as 1, and the lower trap as 2. Gray vertical lines represent the optical pulses driving the transitions between the clock states (internal degrees of freedom). The curly brackets indicate the time under the influence of a certain unitary transformation. The opacity and color of the shaded circles indicate the probability to find an atom at a particular spatial mode. $U_{phase}(T)$ marks the evolution for a duration $T$ of the trapped atoms while they are subjected to the earth's gravitational potential. We depict above the interferometer diagram vectors on the Bloch sphere of  the atomic states in the spin-$\frac{1}{2}$-like space spanned by the clock states at each corresponding time. The blue (red) arrows show the state at the upper (lower) arm.}	\label{fig:graphical_abstract}
\end{figure*}

Our guided ACIF scheme is depicted in Fig.~\ref{fig:graphical_abstract}. It is a modification of the scheme discussed in section \ref{sec:Tweezer atomic interferometry}, and can be applied with both the two- and three-tweezer approaches. Unlike the original tweezer AIF, it employs an atom which can be used as an atomic clock, i.e., with two long-lived internal states, denoted by $\ket{g}$ and $\ket{e}$. We choose the tweezers to operate at the magic wavelength of the clock, where both clock states experience the same trap potential. The atom is initially prepared in the ground state $\ket{g}$. Before the wavepacket splitting stage, a $\pi/2$-pulse of an optical field, which is resonant with the $\ket{g}\leftrightarrow \ket{e}$ transition, is applied. This pulse generates the superposition $\frac{1}{\sqrt{2}}(\ket{g} + \ket{e})$.

Following the $\pi/2$-pulse, the spatial splitting of the clock wavefunction occurs using either two or three tweezers. As detailed in section \ref{sec:Tweezer atomic interferometry}, achieving balanced splitting relies on adiabatic following to reach the symmetric ground state of a symmetric double-well potential. However, in the modified scheme the splitting occurs in the direction defined by gravity. The presence of a gravitational potential disrupts this spatial symmetry, potentially leading to imbalanced splitting. Numerical simulations indicate that Earth's gravitational potential results in a probability distribution of 44\% and 56\% for atoms being in the higher and lower arms, respectively, thus diminishing the visibility of interference fringes. To counteract this, applying a magnetic field gradient to negate the gravitational force during splitting and recombination is effective. Altering the depth of the tweezers during the operation can also compensate for the potential imbalance. Alternatively, conducting the splitting and recombining processes in a plane perpendicular to gravity and subsequently repositioning the arms at different heights can eliminate gravity's influence. For the remainder of this analysis, we assume the imbalance in splitting is not significant and proceed with the evaluation using balanced atomic beam-splitters.

Each interferometer arm is moved to a different position in the gravitational field. Therefore, each part of the clock wavefunction experiences a different proper time. Before the recombining phase, a $\pi$-pulse on the $\ket{g}$-$\ket{e}$ transition is applied to only one of the arms. This pulse is necessary for the interferometer fringe visibility to scale linearly with the redshift, which is essential for making it detectable in a realistic experiment. Since the transitions between the clock states are driven by an optical field, it is straightforward to overlap the driving beam with only one of the tweezers to accomplish this step. The sequence ends with the spatial recombining of the wavefunction, as described in section \ref{sec:Tweezer atomic interferometry}, followed by another $\pi/2$-pulse on the clock degrees of freedom to close the Ramsey sequence. The measured observables are the internal state of the atom and the spatial output port. As we demonstrate below, both quantities are required for a measurement of proper time differences in a coherent superposition of an atom.

To proceed with the calculation, we use a unitary time-evolution operator approach. We label the vectors spanning the relevant Hilbert space by $\ket{k;l}$, where $k=g,e$ and $l=1,2$ denote the internal (clock) and external (paths) degrees of freedom, respectively. The evolution operators employed in the description of the interferometer are defined with respect to the following vectors:
\begin{align}
\begin{split}
\ket{g;1} &\equiv \begin{pmatrix}1, & 0, & 0, & 0\end{pmatrix}\\
\ket{g;2} &\equiv \begin{pmatrix}0, & 1, & 0, & 0\end{pmatrix}\\
\ket{e;1} &\equiv \begin{pmatrix}0, & 0, & 1, & 0\end{pmatrix}\\
\ket{e;2} &\equiv \begin{pmatrix}0, & 0, & 0, & 1\end{pmatrix} \, .
\end{split}
\end{align}
After the splitting and before the recombining, we define the energy difference $\Delta E$ (and the corresponding angular frequency $\omega_{0}$) between the two clock states in path 2 as 
\begin{equation}\label{eq:delta_E_2}
    \Delta E=\hbar\omega_{0}= E_{e;2} - E_{g;2} \, .
\end{equation}
We assume that there is a gravitational potential difference between the paths, $\Delta \varphi$, which gives rise to a difference in the proper time. This means that clocks following the different paths tick at a different rate. Therefore, the transition energy between the clock states must depend on the path, since it can be used as a clock. We write this explicitly as
\begin{equation}\label{eq:delta_E_1}
    E_{e;1} - E_{g;1} = \Delta E + \hbar\epsilon
\end{equation}
with $\epsilon = \frac{\omega_{0}}{c^2}\Delta \varphi$ being the angular frequency difference due to the gravitational redshift. Our goal for the ACIF sequence is to allow determination of $\epsilon$. It is important to note that in the way we wrote Eqs.(\ref{eq:delta_E_2}-\ref{eq:delta_E_1}) we explicitly assumed that the energy difference between the clock states does not depend on the depth of the confining optical potential, which is justified when the trap is operated at the magic wavelength.

To calculate the output of interferometric sequence, we define the unitary operations from which it is composed. $U_{BS}$ is the unitary time-evolution operator corresponding to the spatial wavefunction beam-splitter,
\begin{equation}
    U_{BS} = \frac{1}{\sqrt{2}}\begin{pmatrix}1 & 1 & 0 & 0 \\
   1 & -1 & 0 & 0\\
   0 & 0 & 1 & 1\\
   0 & 0 & 1 & -1
   \end{pmatrix}\, .
\end{equation}
$U_{\pi/2}$ corresponds to a $\pi/2$-pulse over the internal degree of freedom for \emph{both} interferometer arms, 
\begin{equation}
    U_{\pi/2} = \frac{1}{\sqrt{2}}\begin{pmatrix}1 & 0 & -1 & 0 \\
   0 & 1 & 0 & -1\\
   1 & 0 & 1 & 0\\
   0 & 1 & 0 & 1
   \end{pmatrix} \, .
\end{equation}
$U_{\pi,l}$ represents a $\pi$-pulse over the internal degree of freedom in the \emph{lower arm only}, the upper arm remains unchanged,
\begin{equation}
    U_{\pi,l} = \frac{1}{\sqrt{2}}\begin{pmatrix}1 & 0 & 0 & 0 \\
   0 & 0 & 0 & -1\\
   0 & 0 & 1 & 0\\
   0 & 1 & 0 & 0
   \end{pmatrix} \, .
\end{equation}
$U_{\text{phase}}(T)$ corresponds to the phase accumulation stage with a duration $T$,
\begin{equation}\label{Eq_U_phase}
    U_{\text{phase}}(T) =
    \begin{pmatrix}1 & 0 & 0 & 0 \\
   0 & e^{-i\delta T} & 0 & 0\\
   0 & 0 & e^{i(\Delta +\epsilon) T} & 0\\
   0 & 0 & 0 & e^{i(\Delta-\delta) T}
   \end{pmatrix} \, ,
\end{equation}
where we omit a global phase and define the detuning between the two lower state, $\delta=(E_{g;2} - E_{g;1})/\hbar$, and the drive detuning, $\Delta=\omega-\omega_0$, with $\omega$ being the frequency of the field driving the transitions involving the internal states. Eq. \ref{Eq_U_phase} is written in a rotating frame defined by the transformation
\begin{equation}\label{Eq_U_rotating_frame}
    U_{\text{rot}}(T) =
    \begin{pmatrix} 1 & 0 & 0 & 0 \\
   0 & 1 & 0 & 0\\
   0 & 0 & e^{i\omega T} & 0\\
   0 & 0 & 0 & e^{i\omega T}
   \end{pmatrix} \, .
\end{equation}
The final state after our proposed ACIF sequence can be written as
\begin{equation}
    \ket{\psi_{f}} = U_{\text{rot}}^\dagger (T)U_{\pi/2}^{\dagger}U_{BS}^{\dagger}U_{\pi,l}U_{\text{phase}}(T)U_{BS}U_{\pi/2} \ket{g;1} \, .
\end{equation}
Note that we omit an initial $U_{\text{rot}}(0)$ since it is an identity.

In previous proposals of ACIF schemes, there is only one available observable for measurements of the interference pattern of the atomic phase due to proper time differences between the arms - either the spatial output port \cite{Rosi_2017,Loriani_2019,Roura_2020}, or the internal clock state \cite{Di_Pumpo_2021}. A unique feature of the proposed tweezer ACIF is that it has two spatial output ports and complete Ramsey sequence over the clock states, both exhibit coherent oscillations. The probability of the atoms to exit the interferometer from the upper port is given by
\begin{align}\label{eq:p_output}
\begin{split}
    &P_{1}(\psi_{f}) = P_{g;1} + P_{e;1}
    = |\langle g;1|\psi_{f}\rangle|^{2} +|\langle e;1|\psi_{f}\rangle|^{2}= \\
    &=\frac{1}{2}\Biggl[1-\sin\Bigl[T\bigl(\frac{E_{g;2}-E_{g;1}}{\hbar}-\frac{\epsilon}{2}\bigr)\Bigr]\sin\Bigl[T\bigl(\Delta-\frac{\epsilon}{2}\bigr)\Bigr]\Biggr] \, .
\end{split}
\end{align}
A crucial requirement for an interferometric measurement is the ability to verify that the atom was indeed in a coherent superposition during the sequence. The result of Eq.(\ref{eq:p_output}) allows this, since it exhibits coherent oscillations between the exit ports as a function of $T$. In contrast, if the wavepacket collapses randomly to one of the paths and the state becomes completely mixed, the exit port probability is $P_{1,\text{mixed}}(\psi_{f})=\frac{1}{2}$.  This means that measuring oscillations around the probability of $\frac{1}{2}$ as a function of time verifies the coherence of the atomic wavefunction.

The primary drawback of measuring the spatial output port is its exit probability's reliance on the difference between the eigenenergies: $E_{g;1}-E_{g;2}$. This dependency makes the observable vulnerable to relative intensity fluctuations of the trap beams. To address this issue, there are two strategies. Firstly, one could impose strict requirements on laser intensity stability throughout the interferometer's duration to ensure minimal noise in this measurement. In conventional atomic interferometry, this is the sole approach since one needs to determine the relative phase between the paths, which is done by measuring the phase of the exit port oscillations. 

In our case, however, we only need to verify the coherence of the split wavepacket, which allows for a simpler approach. Intensity differences between the tweezers (intentional or unintentional) can introduce essentially a random phase in the first sine term of Eq.(\ref{eq:p_output}). Given $x=\frac{1}{2}-A\sin(\phi)$ with $A\in [0,0.5]$ and the phase $\phi$ having a uniform random distribution in $[0,2\pi]$, the probability distribution of $x \in [0.5-A,0.5+A]$ is ${P(x)=\frac{1}{\pi\sqrt{A^2-(0.5-x)^2}}}$. In the case of Eq.(\ref{eq:p_output}), ${A=\frac{1}{2}\sin\Bigl[T\bigl(\Delta-\frac{\epsilon}{2}\bigr)\Bigr]\Biggr]}$ is set by the choice of $T$ and $\Delta$ while $\epsilon$ can be neglected. $T$ and $\Delta$ are well controlled in the experiment, and therefore $A$ can be regarded as constant. Thus, if the wavepacket is coherently split and maintained, the distribution of the estimator of $P_{1}$ should align with $P(x)$ across a sufficiently large dataset. Since $P_1$ changes from run to run, it is necessary to estimate it with sufficient accuracy in each run. This requires interferometeric runs with several atoms. As we show in the next section, 10 atoms per run are already enough to generate a clear distinction between coherent and non-coherent wavepacket. 

With this approach we are still able to measure the gravitational redshift, $\epsilon$, thanks to to the availability of an additional observable - the probability of the atoms to be in the clock ground state at the end of the sequence (regardless of the output port). It is given by
\begin{align}\label{eq:main_result}
\begin{split}
    P_{g}(\psi_f) &= P_{g;1} + P_{g;2}
    = |\langle g;1|\psi_{f}\rangle|^{2} +|\langle g;2|\psi_{f}\rangle|^{2} \\
    &=\frac{1}{2}\Bigg[1+\sin\Bigl[T\bigl(\Delta-\frac{\epsilon}{2}\bigr)\Bigr]\sin\Bigl[T\frac{\epsilon}{2}\Bigr]\Biggr] \, .
\end{split}
\end{align}
As is readily seen from Eq.(\ref{eq:main_result}), in contrast to the case of measuring the spatial output port, the probability to finish in the ground state depends \emph{only} on $\Delta$ and $\epsilon$, but not on the eigenenergies separately. Thanks to that, this observable is insensitive to relative fluctuations in the depths of the tweezers, which would shift the eigenenergies, but do not affect $\omega_{0}$ and $\Delta$, as long as we operate at the magic wavelength. This means that for measuring the internal clock state, the main noise source limiting guided AIFs is highly suppressed.

The probability to be in the ground state is the same whether the state is coherent or mixed. Therefore, it is essential to measure both the spatial output port and the internal state of the atom. The measurement of the output port will ensure the coherence of the atomic state, while the measurement of the internal state, which is robust in the presence of laser intensity noise, will be used to extract $\epsilon$.

The redshift is extracted from Eq.(\ref{eq:main_result}). By scanning the waiting time, $T$, the probability to find the atoms in the ground state oscillates due to interference of the two paths. The oscillation frequency of this interference pattern is set by the detuning of the driving field relative to the clock frequency, $\Delta$, shifted by $\frac{\epsilon}{2}$. Choosing ${\Delta \gg \frac{2\pi}{T},\epsilon}$ ensures that the oscillations can be observed by scanning the time from $T$ to $T+2\pi/\Delta$, during which the amplitude of the oscillations is almost constant and given by
\begin{equation}\label{eq:visibility_eq}
    \mathcal{V} =  \sin(T\frac{\epsilon}{2}) \, .
\end{equation}
Since this amplitude is determined by $\epsilon$, it gives a straightforward way to extract its value. Importantly, for small redshifts ($T\epsilon\ll 1$), the visibility scales favorably \emph{linearly} with $\epsilon$: $\mathcal{V} \approx T\epsilon/2$. This should be contrasted with a similar interferometric sequence, only without the $\pi$ pulse in the lower path. In that case, the visibility of the interference pattern scales as $\mathcal{V} \approx 1-(T\epsilon)^2/8$. The linear scaling of the visibility in our scheme is crucial to make an experimental observation feasible, as we illustrate below.

\section{Feasibility analysis}\label{sec:Feasibility Analysis}
We first estimate the measurable effect in a realistic experimental scenario. The redshift is 
\begin{equation}
	\epsilon = 
	 \frac{\omega_{0}}{c^2}g h \, \, ,
\end{equation}	
with $ h$ being the height difference between the two interferometer arms. We take the atom to be $^{171}$Yb, with the clock transition being ${}^{1}S_0-{}^{3}P_0$. The energy difference between the clock states corresponds to optical emission with a wavelength of approximately $\lambda=578$nm. The magic wavelength of the tweezer trap for this atom is around $\lambda_{\text{magic}} = 759$nm \cite{Hinkley_2013}. 

The resilience of the measurement to differential intensity fluctuations depends on the stability of the tweezer wavelength. Deviations from the magic wavelength result in intensity-dependent light-shifts in the clock states frequencies \cite{Katori_2015, Madjarov_2019}, which in turn may reduce the accuracy. The tweezers's wavelength can be stabilized to the desired value with sub-MHz resolution \cite{Hinkley_2013,Okuno_2022}, resulting in frequency noise in the order of $10^{-5}$Hz. This noise level has no effect on the measurement accuracy, as we have confirmed through numerical simulations.

To minimize motional transitions in the tweezer during the optical pulse, it is required to operate within the Lamb-Dicke regime \cite{Ludlow_2015}. This regime is characterized by a small $\eta = \frac{2\pi}{\lambda}\cdot x_0$, where $x_0$ represents the spatial extent of the atomic wavefunction trapped in the tweezer. For a tweezer with a depth of $300\mu$K and a waist of $1\mu$m, the resulting Lamb-Dicke parameters are approximately $\eta \approx 0.3$ and $0.73$ in the radial and axial directions, respectively. These values fall within the operational range for optical atomic clocks \cite{Young_2020}, suggesting they should be adequate for the interferometer. During the phase accumulation stage, in the absence of optical pulses, a deep trap is unnecessary. To lower the chance of spontaneous emission from the tweezer light during this extended phase, the depth of the tweezers should be reduced and then increased again just before the recombination stage.

Similar to Ref. \cite{Nemirovsky_2023}, we assume a separation between the two tweezer arms of $h=10$ mm, aligned in the same direction as Earth's gravitational acceleration, $g$. Taking a phase accumulation duration of $T=10s$, we obtain a visibility of $\mathcal{V} \approx 0.02$. The oscillating signal appears on top of a background signal of 0.5. To estimate the number of runs required to clearly observe the gravitational effect, we have performed a Monte Carlo simulation of the entire experiment. In the simulation, we set the detuning to $\Delta = 2\pi \cdot 1000$ Hz and scan $N_1$ different durations in the range $T \in [10, 10 + \frac{2\pi}{\Delta}]$. For each of these durations, we simulate the measurement with $N_a$ atoms according to Eqs. \ref{eq:p_output} and \ref{eq:main_result}. We then repeat the procedure $N_2$ times for each duration to get an estimator of $P_g(T)$ and $P_1(T)$. We fit the fringe of the former, and from the extracted visibility, we find $\epsilon$. Additionally, we included intensity fluctuations between the tweezer arms by adding a random energy shift on the order of $\hbar \Delta$ between the $E_{g;2}$ and $E_{g;1}$ states.

\begin{figure}
    \centering
    \includegraphics[width=1\linewidth]{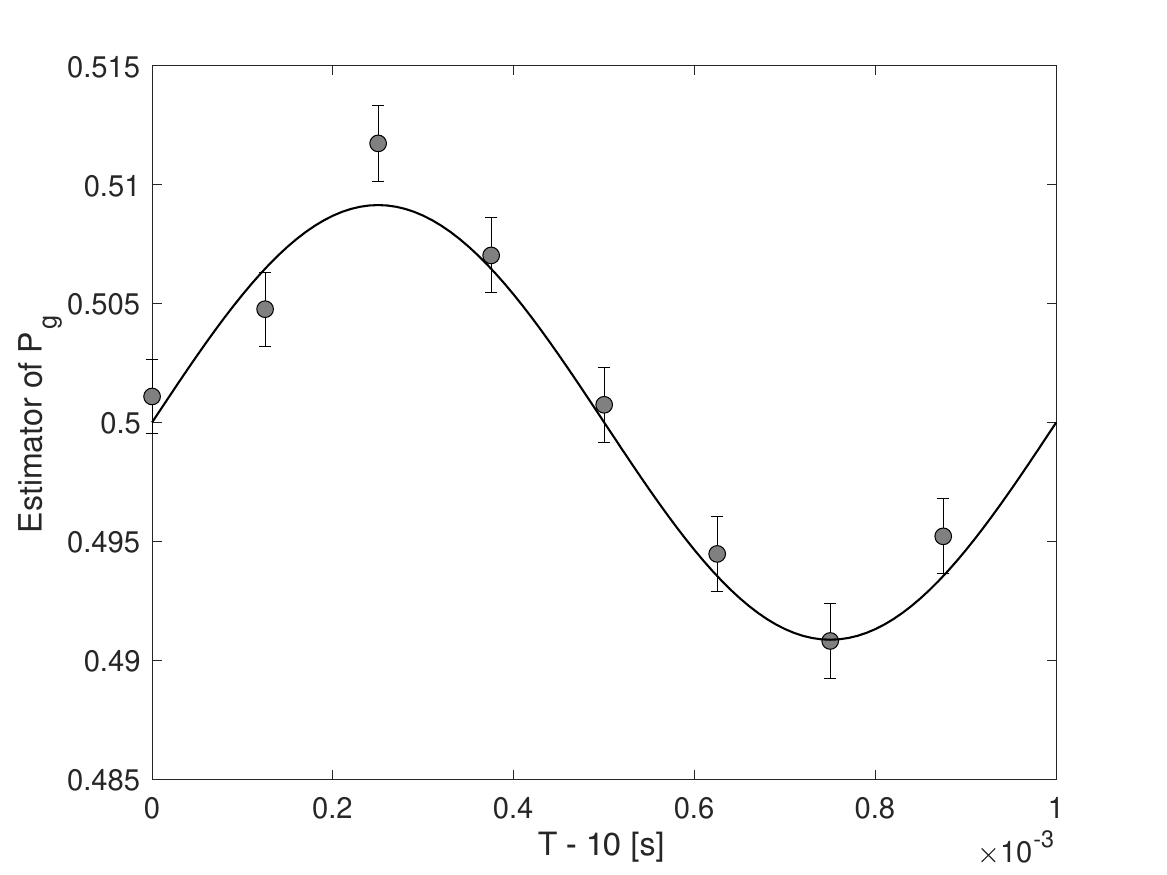}
    \caption{\textbf{Numerical simulation of the interferometric measurement with a separation of $h=10$ mm in Earth’s gravitational field.} In this simulation, we used $\Delta=2\pi \cdot 1000$ Hz, $N_a=20$ atoms per run, and $N_2=5000$ repetitions of each duration. The circular markers represent the average of the estimator of $P_g$, with error bars that represent the standard error. The solid line is the fit to Equation \ref{eq:main_result}, with the amplitude and phase being the free parameters.}
    \label{fig:Monte_Carlo_single_fringe}
\end{figure}

A typical result of the simulation is shown in figure \ref{fig:Monte_Carlo_single_fringe}. In this example, we have taken $N_a=20$ atoms per run and $N_2=5000$ repetitions. Factoring in an overhead of 5 seconds for each run, the total duration per run extends to 15 seconds. Scanning the fringe at $N_1=8$ different phases, as shown in the figure, requires about a week of data collection. With these parameters, the fringe contrast is clearly visible, and the relative accuracy (one standard deviation) in determining $\epsilon$ is approximately $9\%$. Importantly, the accuracy of the extracted $\epsilon$ is not compromised by the intensity fluctuations between the tweezers. In Table \ref{tab:exp_parameters}, we present the relative accuracy for other realistic choices of parameters. In all cases, the accuracy allows unambiguous determination of a gravitational redshift effect.

\begin{table}
	\begin{tabular}{|>{\centering\arraybackslash}p{1.3cm}|>{\centering\arraybackslash}p{2cm}|>{\centering\arraybackslash}p{1.3cm}|>{\centering\arraybackslash}p{1.7cm}|>{\centering\arraybackslash}p{1.4cm}|}
		\hline
		Number of atoms in a run ($N_a$) & Number of repetitions per phase and duration ($N_2$) & Probing duration (T) [s] & total runtime & Relative accuracy in extracting $\epsilon$  \\
		\hline \hline
            100 & 5000 & 1  &  $\sim$ 2.8 days & 38.1\% \\ 
            \hline
            100 & 10000 & 1  &  $\sim$ 5.6 days & 28.4\% \\ 
            \hline
            20 	& 5000 & 3  &  $\sim$ 3.7 days & 28.9\% \\
            \hline
            100 & 5000 & 3  &  $\sim$ 3.7 days & 12.9\% \\
            \hline
            100 & 10000 & 3  &  $\sim$ 7.4 days & 9.8\% \\
		\hline
            10 	& 1000 & 10  &  $\sim$ 1.4 days & 28.4\% \\
		\hline
            10 	& 5000 & 10  &  $\sim$ 7 days & 12.4\% \\
		\hline
		20 	& 5000 & 10  &  $\sim$ 7 days & 8.8\% \\ 
            \hline
            100 & 1000 & 10  &  $\sim$ 1.4 days & 9\% \\ 
            \hline
            100 & 5000 & 10  &  $\sim$ 7 days & 3.8\% \\ 
		\hline
            100 & 10000 & 10  &  $\sim$ 13.9 days & 2.7\% \\ 
            \hline
	\end{tabular}
	\caption{\textbf{Different possible choices of experimental parameters.} In all cases, we assume a separation of $h=10$ mm between the interferometer arms in Earth’s gravitational field. Similar to figure \ref{fig:Monte_Carlo_single_fringe}, we fix the number of different phases to $N_1=8$. The total runtime takes into account an overhead of $5$ seconds per run. The relative accuracy is defined as one standard deviation of the extracted values of $\epsilon$ over an ensemble of $1000$ Monte Carlo simulation runs, normalized by the theoretical value of $\epsilon$.} \label{tab:exp_parameters}
\end{table}

\begin{figure}[ht]
	\centering
	\includegraphics[width=\linewidth]{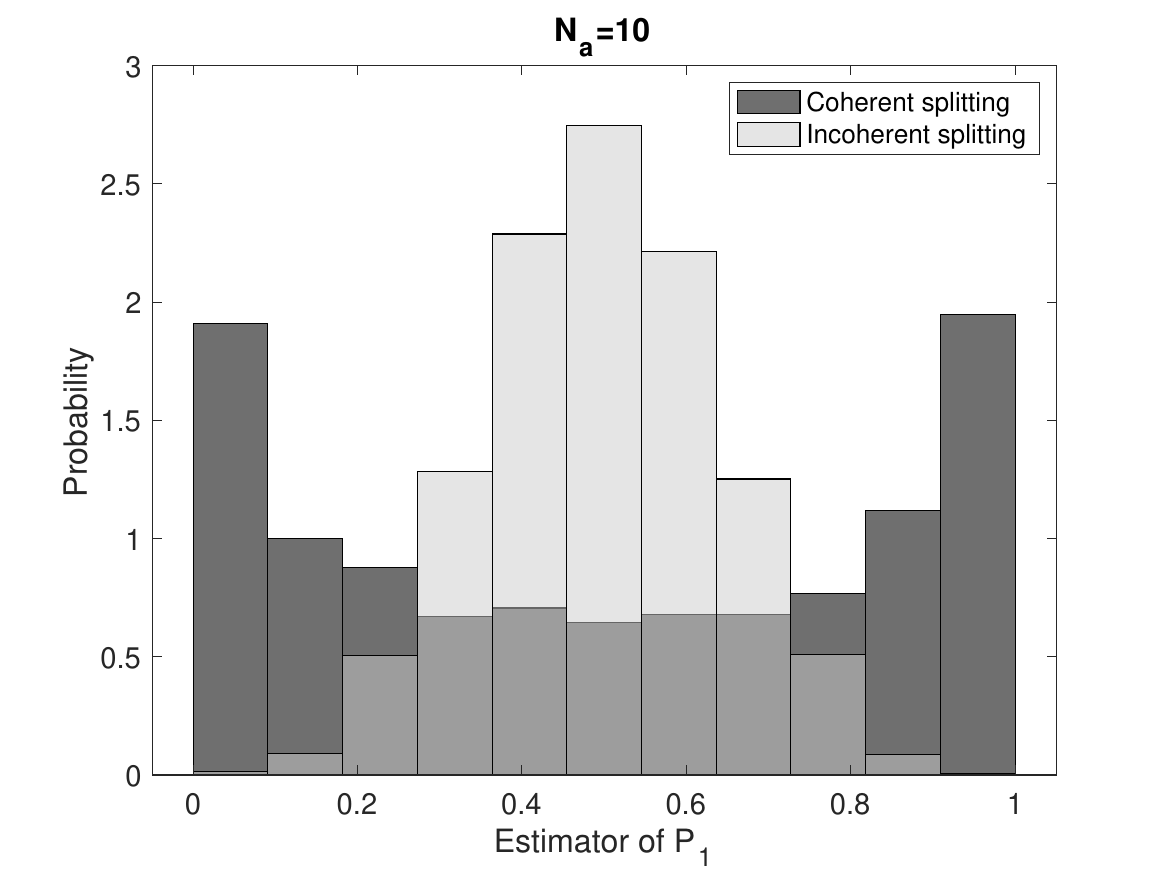}
	\includegraphics[width=\linewidth]{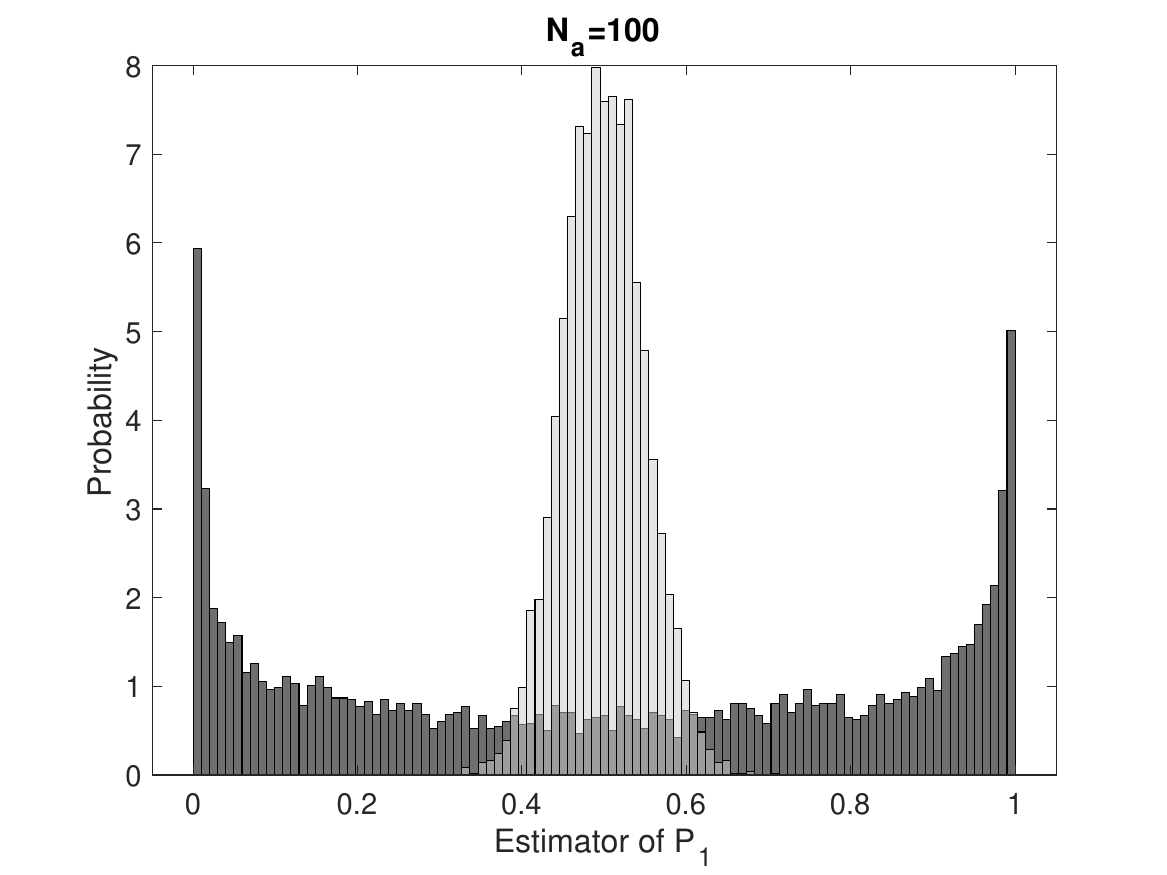}
	\caption{\textbf{Histograms of the estimator of $P_1$ for coherent and incoherent splitting.} The histograms are from an ensemble of 1000 Monte Carlo simulation runs that include strong intensity fluctuations between the tweezer arms. In the incoherent splitting case (brighter shading), we assume that the wavepacket randomly collapses to one of the arms after the splitting stage. For both $N_a=10$ atoms (top) and $N_a=100$ atoms (bottom) per run, the coherent and incoherent histograms are easily distinguished. The parameters in these simulations are $T=10.00025$ sec, $\Delta=2\pi\cdot 1000$Hz, $h=10$mm, and $N_2=5000$. The different discretization of the x-axis is due to the different $N_a$, which results in a different finite set of possible values of $P_1$ in each run.}
	\label{fig:P1_histograms}
\end{figure}

The Monte Carlo simulations allow us to test our approach to verifying the coherent splitting. The random fluctuations we introduce in the relative ground state energies translate into essentially a random phase of the first sine term in Eq. \ref{eq:p_output}. In figure \ref{fig:P1_histograms}, we plot the histogram of the estimator of $P_1$ (darker shading) for $10$ (top) and $100$ (bottom) atoms in each run. For comparison, we also depict the histogram obtained if a collapse randomly occurs to one of the two arms (brighter shading). The coherent and incoherent histograms are markedly different. In particular, the probability of finding highly unbalanced splitting between the output ports is negligibly small when the wavepacket decoheres, while it is maximal if it maintains its coherence. These results prove that the coherence of the wavepacket can be verified even in the presence of strong intensity fluctuations and with a relatively small number of atoms.

Operating two optical Ramsey-like pulses coherently over several seconds requires a narrow linewidth of the clock laser. This requirement is similar to what is needed in light-pulse interferometers using the optical transition in ${}^{88}$Sr \cite{Chiarotti_2022}. State-of-the-art laser systems built for atomic clocks can achieve a linewidth below 10 mHz \cite{Matei2017,Oelker_2019}. A recent experiment with a tweezer atomic clock reported an atom-light coherence time of 3.6 seconds \cite{Young_2020}. The reduced coherence time relative to the limit dictated by the laser linewidth is likely due to fluctuations in the tweezer position, or uncontrolled phase noise from environmental factors. Even so, as shown in Table \ref{tab:exp_parameters}, this time is sufficient to determine the redshift with an accuracy better than $10\%$. We conclude that the observation of gravitational redshift with ACIF is within the reach of current technological capabilities.

\section{Discussion}\label{sec:Discussion}

We introduce a guided atomic clock interferometer approach, an extension of our previous tweezer interferometer method \cite{Nemirovsky_2023}, tailored for atoms with two internal states. This technique employs optical laser pulses to create a superposition of the internal states and utilizes precise manipulation of the tweezers' position and intensity for the spatial adiabatic splitting and recombining of the wavepacket. After completion of the interferometric scheme, we record the population within each clock state and exit port. The statistical distribution of the splitting between the two exit ports confirms the wavepacket's spatial coherence throughout the experiment. Oscillations between the internal states reveal time dilation across the paths. In particular, the visibility of these oscillations is instrumental in determining the gravitational redshift between the interferometer arms. Clock interferometry, which has not been performed to date, is achievable using our proposed scheme with the current state of technology.

Quantum theory has been tested successfully in many experiments, including with atomic interferometry. Similarly, predictions of General Relativity have also been verified, including measurements of the gravitational redshift using two separate clocks \cite{Vessot_1980,Takamoto_2020,Chou_2010,Bothwell_2022,Herrmann_2018,Delva_2018}. However, theories concerning  regimes in which both quantum mechanics and general relativity are relevant remain untested. The proposed ACIF could probe this unexplored regime for the first time \cite{Zych_2016}. 

One theory on the intersection of quantum mechanics and general relativity suggests that gravitational redshift contributes to the decoherence observed in the classical limit \cite{Pikovski_2015}. In the context of a clock interferometer, the gravitational field causes entanglement between the atom's internal state and its spatial wavefunction. In a larger scale scenario, a body composed of many such internal degrees of freedom can be viewed as multiple clocks operating at varying rates, influenced by the gravitational field. This variance in ticking rates leads to dephasing among the clocks and consequently reduction in coherence for the spatial wavefunction. This offers a potential explanation for decoherence in the classical limit that does not rely on interaction with environment. Using an ACIF to test the case of a single clock is a first step towards testing the effect of gravity on the coherence of macroscopic objects.

An ACIF experiment may also have implications for theories of quantum time. In general relativity, time is dynamic and dependent on the metric, functioning as a dimension. Conversely, quantum theory treats time as a global parameter. To resolve this discrepancy between general relativity and quantum theory, there have been suggestions for a dynamic quantum time operator \cite{Paiva_2022,Giovannetti_2015}. Some of these suggestions advocate for considering proper time as the quantum operator with mass being its conjugate. This allows for a quantum operator representation of mass as well \cite{Greenberger1970,Greenberger_1970}. If proper time is represented by a quantum operator, its uncertainty could influence the expected visibility of an ACIF experiment. Therefore, measuring deviations from the expected result could help constrain theories of quantum proper time \cite{Zych_2011}.

\begin{acknowledgments}
We thank Jonathan Nemirovsky, Eliahu Cohen, and Ran Finkelstein for helpful discussions. We thank Anastasiya Vainbaum, Amir Stern, and Magdalena Zych for critical reading of the manuscript. This research was supported by the Israel Science Foundation (ISF), grant No. 3491/21, and by the Pazy Research Foundation. This research project was partially supported by the Helen Diller Quantum Center at the Technion.
\end{acknowledgments}



\begin{thebibliography}{59}%
	\makeatletter
	\providecommand \@ifxundefined [1]{%
		\@ifx{#1\undefined}
	}%
	\providecommand \@ifnum [1]{%
		\ifnum #1\expandafter \@firstoftwo
		\else \expandafter \@secondoftwo
		\fi
	}%
	\providecommand \@ifx [1]{%
		\ifx #1\expandafter \@firstoftwo
		\else \expandafter \@secondoftwo
		\fi
	}%
	\providecommand \natexlab [1]{#1}%
	\providecommand \enquote  [1]{``#1''}%
	\providecommand \bibnamefont  [1]{#1}%
	\providecommand \bibfnamefont [1]{#1}%
	\providecommand \citenamefont [1]{#1}%
	\providecommand \href@noop [0]{\@secondoftwo}%
	\providecommand \href [0]{\begingroup \@sanitize@url \@href}%
	\providecommand \@href[1]{\@@startlink{#1}\@@href}%
	\providecommand \@@href[1]{\endgroup#1\@@endlink}%
	\providecommand \@sanitize@url [0]{\catcode `\\12\catcode `\$12\catcode
		`\&12\catcode `\#12\catcode `\^12\catcode `\_12\catcode `\%12\relax}%
	\providecommand \@@startlink[1]{}%
	\providecommand \@@endlink[0]{}%
	\providecommand \url  [0]{\begingroup\@sanitize@url \@url }%
	\providecommand \@url [1]{\endgroup\@href {#1}{\urlprefix }}%
	\providecommand \urlprefix  [0]{URL }%
	\providecommand \Eprint [0]{\href }%
	\providecommand \doibase [0]{https://doi.org/}%
	\providecommand \selectlanguage [0]{\@gobble}%
	\providecommand \bibinfo  [0]{\@secondoftwo}%
	\providecommand \bibfield  [0]{\@secondoftwo}%
	\providecommand \translation [1]{[#1]}%
	\providecommand \BibitemOpen [0]{}%
	\providecommand \bibitemStop [0]{}%
	\providecommand \bibitemNoStop [0]{.\EOS\space}%
	\providecommand \EOS [0]{\spacefactor3000\relax}%
	\providecommand \BibitemShut  [1]{\csname bibitem#1\endcsname}%
	\let\auto@bib@innerbib\@empty
	\bibitem [{\citenamefont {Cronin}\ \emph {et~al.}(2009)\citenamefont {Cronin},
		\citenamefont {Schmiedmayer},\ and\ \citenamefont {Pritchard}}]{Cronin_2009}%
	\BibitemOpen
	\bibfield  {author} {\bibinfo {author} {\bibfnamefont {A.~D.}\ \bibnamefont
			{Cronin}}, \bibinfo {author} {\bibfnamefont {J.}~\bibnamefont
			{Schmiedmayer}},\ and\ \bibinfo {author} {\bibfnamefont {D.~E.}\ \bibnamefont
			{Pritchard}},\ }\bibfield  {title} {\bibinfo {title} {Optics and
			interferometry with atoms and molecules},\ }\href
	{https://doi.org/10.1103/RevModPhys.81.1051} {\bibfield  {journal} {\bibinfo
			{journal} {Reviews of Modern Physics}\ }\textbf {\bibinfo {volume} {81}},\
		\bibinfo {pages} {1051} (\bibinfo {year} {2009})}\BibitemShut {NoStop}%
	\bibitem [{\citenamefont {Rafi~Weill}\ and\ \citenamefont
		{Sagi}(2023)}]{Nemirovsky_2023}%
	\BibitemOpen
	\bibfield  {author} {\bibinfo {author} {\bibfnamefont {J.~N.}\ \bibnamefont
			{Rafi~Weill}, \bibfnamefont {Ilan~Meltzer}}\ and\ \bibinfo {author}
		{\bibfnamefont {Y.}~\bibnamefont {Sagi}},\ }\bibfield  {title} {\bibinfo
		{title} {Atomic interferometer based on optical tweezers},\ }\href
	{https://doi.org/10.1103/PhysRevResearch.5.043300} {\bibfield  {journal}
		{\bibinfo  {journal} {Physical Review Research}\ }\textbf {\bibinfo {volume}
			{5}},\ \bibinfo {pages} {043300} (\bibinfo {year} {2023})}\BibitemShut
	{NoStop}%
	\bibitem [{\citenamefont {Rasel}\ \emph {et~al.}(1995)\citenamefont {Rasel},
		\citenamefont {Oberthaler}, \citenamefont {Batelaan}, \citenamefont
		{Schmiedmayer},\ and\ \citenamefont {Zeilinger}}]{Rasel_1995}%
	\BibitemOpen
	\bibfield  {author} {\bibinfo {author} {\bibfnamefont {E.~M.}\ \bibnamefont
			{Rasel}}, \bibinfo {author} {\bibfnamefont {M.~K.}\ \bibnamefont
			{Oberthaler}}, \bibinfo {author} {\bibfnamefont {H.}~\bibnamefont
			{Batelaan}}, \bibinfo {author} {\bibfnamefont {J.}~\bibnamefont
			{Schmiedmayer}},\ and\ \bibinfo {author} {\bibfnamefont {A.}~\bibnamefont
			{Zeilinger}},\ }\bibfield  {title} {\bibinfo {title} {Atom wave
			interferometry with diffraction gratings of light},\ }\href
	{https://doi.org/10.1103/PhysRevLett.75.2633} {\bibfield  {journal} {\bibinfo
			{journal} {Physical Review Letters}\ }\textbf {\bibinfo {volume} {75}},\
		\bibinfo {pages} {2633} (\bibinfo {year} {1995})}\BibitemShut {NoStop}%
	\bibitem [{\citenamefont {Giltner}\ \emph {et~al.}(1995)\citenamefont
		{Giltner}, \citenamefont {McGowan},\ and\ \citenamefont
		{Lee}}]{Giltner_1995}%
	\BibitemOpen
	\bibfield  {author} {\bibinfo {author} {\bibfnamefont {D.~M.}\ \bibnamefont
			{Giltner}}, \bibinfo {author} {\bibfnamefont {R.~W.}\ \bibnamefont
			{McGowan}},\ and\ \bibinfo {author} {\bibfnamefont {S.~A.}\ \bibnamefont
			{Lee}},\ }\bibfield  {title} {\bibinfo {title} {Atom interferometer based on
			bragg scattering from standing light waves},\ }\href
	{https://doi.org/10.1103/PhysRevLett.75.2638} {\bibfield  {journal} {\bibinfo
			{journal} {Physical Review Letters}\ }\textbf {\bibinfo {volume} {75}},\
		\bibinfo {pages} {2638} (\bibinfo {year} {1995})}\BibitemShut {NoStop}%
	\bibitem [{\citenamefont {Raithel}\ \emph {et~al.}(2022)\citenamefont
		{Raithel}, \citenamefont {Duspayev}, \citenamefont {Dash}, \citenamefont
		{Carrasco}, \citenamefont {Goerz}, \citenamefont {Vuletić},\ and\
		\citenamefont {Malinovsky}}]{Raithel_2022}%
	\BibitemOpen
	\bibfield  {author} {\bibinfo {author} {\bibfnamefont {G.}~\bibnamefont
			{Raithel}}, \bibinfo {author} {\bibfnamefont {A.}~\bibnamefont {Duspayev}},
		\bibinfo {author} {\bibfnamefont {B.}~\bibnamefont {Dash}}, \bibinfo {author}
		{\bibfnamefont {S.~C.}\ \bibnamefont {Carrasco}}, \bibinfo {author}
		{\bibfnamefont {M.~H.}\ \bibnamefont {Goerz}}, \bibinfo {author}
		{\bibfnamefont {V.}~\bibnamefont {Vuletić}},\ and\ \bibinfo {author}
		{\bibfnamefont {V.~S.}\ \bibnamefont {Malinovsky}},\ }\bibfield  {title}
	{\bibinfo {title} {Principles of tractor atom interferometry},\ }\href
	{https://doi.org/10.1088/2058-9565/ac9429} {\bibfield  {journal} {\bibinfo
			{journal} {Quantum Science and Technology}\ }\textbf {\bibinfo {volume}
			{8}},\ \bibinfo {pages} {014001} (\bibinfo {year} {2022})}\BibitemShut
	{NoStop}%
	\bibitem [{\citenamefont {Kasevich}\ and\ \citenamefont
		{Chu}(1991)}]{Kasevich_1991}%
	\BibitemOpen
	\bibfield  {author} {\bibinfo {author} {\bibfnamefont {M.}~\bibnamefont
			{Kasevich}}\ and\ \bibinfo {author} {\bibfnamefont {S.}~\bibnamefont {Chu}},\
	}\bibfield  {title} {\bibinfo {title} {Atomic interferometry using stimulated
			raman transitions},\ }\href {https://doi.org/10.1103/PhysRevLett.67.181}
	{\bibfield  {journal} {\bibinfo  {journal} {Physical Review Letters}\
		}\textbf {\bibinfo {volume} {67}},\ \bibinfo {pages} {181} (\bibinfo {year}
		{1991})}\BibitemShut {NoStop}%
	\bibitem [{\citenamefont {Shin}\ \emph {et~al.}(2004)\citenamefont {Shin},
		\citenamefont {Saba}, \citenamefont {Pasquini}, \citenamefont {Ketterle},
		\citenamefont {Pritchard},\ and\ \citenamefont {Leanhardt}}]{Shin2004}%
	\BibitemOpen
	\bibfield  {author} {\bibinfo {author} {\bibfnamefont {Y.}~\bibnamefont
			{Shin}}, \bibinfo {author} {\bibfnamefont {M.}~\bibnamefont {Saba}}, \bibinfo
		{author} {\bibfnamefont {T.~A.}\ \bibnamefont {Pasquini}}, \bibinfo {author}
		{\bibfnamefont {W.}~\bibnamefont {Ketterle}}, \bibinfo {author}
		{\bibfnamefont {D.~E.}\ \bibnamefont {Pritchard}},\ and\ \bibinfo {author}
		{\bibfnamefont {A.~E.}\ \bibnamefont {Leanhardt}},\ }\bibfield  {title}
	{\bibinfo {title} {Atom interferometry with bose-einstein condensates in a
			double-well potential},\ }\href
	{https://doi.org/10.1103/physrevlett.92.050405} {\bibfield  {journal}
		{\bibinfo  {journal} {Physical Review Letters}\ }\textbf {\bibinfo {volume}
			{92}},\ \bibinfo {pages} {050405} (\bibinfo {year} {2004})}\BibitemShut
	{NoStop}%
	\bibitem [{\citenamefont {Xu}\ \emph {et~al.}(2019)\citenamefont {Xu},
		\citenamefont {Jaffe}, \citenamefont {Panda}, \citenamefont {Kristensen},
		\citenamefont {Clark},\ and\ \citenamefont {Müller}}]{Xu_2019}%
	\BibitemOpen
	\bibfield  {author} {\bibinfo {author} {\bibfnamefont {V.}~\bibnamefont
			{Xu}}, \bibinfo {author} {\bibfnamefont {M.}~\bibnamefont {Jaffe}}, \bibinfo
		{author} {\bibfnamefont {C.~D.}\ \bibnamefont {Panda}}, \bibinfo {author}
		{\bibfnamefont {S.~L.}\ \bibnamefont {Kristensen}}, \bibinfo {author}
		{\bibfnamefont {L.~W.}\ \bibnamefont {Clark}},\ and\ \bibinfo {author}
		{\bibfnamefont {H.}~\bibnamefont {Müller}},\ }\bibfield  {title} {\bibinfo
		{title} {Probing gravity by holding atoms for 20 seconds},\ }\href
	{https://doi.org/10.1126/science.aay6428} {\bibfield  {journal} {\bibinfo
			{journal} {Science}\ }\textbf {\bibinfo {volume} {366}},\ \bibinfo {pages}
		{745} (\bibinfo {year} {2019})}\BibitemShut {NoStop}%
	\bibitem [{\citenamefont {Ryu}\ and\ \citenamefont
		{Boshier}(2015)}]{C_Ryu_2015}%
	\BibitemOpen
	\bibfield  {author} {\bibinfo {author} {\bibfnamefont {C.}~\bibnamefont
			{Ryu}}\ and\ \bibinfo {author} {\bibfnamefont {M.~G.}\ \bibnamefont
			{Boshier}},\ }\bibfield  {title} {\bibinfo {title} {Integrated coherent
			matter wave circuits},\ }\href
	{https://doi.org/10.1088/1367-2630/17/9/092002} {\bibfield  {journal}
		{\bibinfo  {journal} {New Journal of Physics}\ }\textbf {\bibinfo {volume}
			{17}},\ \bibinfo {pages} {092002} (\bibinfo {year} {2015})}\BibitemShut
	{NoStop}%
	\bibitem [{\citenamefont {Gayathrini~Premawardhana}\ and\ \citenamefont
		{Taylor}()}]{GayathriniPremawardhana}%
	\BibitemOpen
	\bibfield  {author} {\bibinfo {author} {\bibfnamefont {S.~S.}\ \bibnamefont
			{Gayathrini~Premawardhana}, \bibfnamefont {Jonathan~Kunjummen}}\ and\
		\bibinfo {author} {\bibfnamefont {J.~M.}\ \bibnamefont {Taylor}},\ }\bibfield
	{title} {\bibinfo {title} {Feasibility of a trapped atom interferometer with
			accelerating optical traps},\ }\href@noop {} {\ }\BibitemShut {NoStop}%
	\bibitem [{\citenamefont {Morel}\ \emph {et~al.}(2020)\citenamefont {Morel},
		\citenamefont {Yao}, \citenamefont {Cladé},\ and\ \citenamefont
		{Guellati-Khélifa}}]{Morel_2020}%
	\BibitemOpen
	\bibfield  {author} {\bibinfo {author} {\bibfnamefont {L.}~\bibnamefont
			{Morel}}, \bibinfo {author} {\bibfnamefont {Z.}~\bibnamefont {Yao}}, \bibinfo
		{author} {\bibfnamefont {P.}~\bibnamefont {Cladé}},\ and\ \bibinfo {author}
		{\bibfnamefont {S.}~\bibnamefont {Guellati-Khélifa}},\ }\bibfield  {title}
	{\bibinfo {title} {Determination of the fine-structure constant with an
			accuracy of 81 parts per trillion},\ }\href
	{https://doi.org/10.1038/s41586-020-2964-7} {\bibfield  {journal} {\bibinfo
			{journal} {Nature}\ }\textbf {\bibinfo {volume} {588}},\ \bibinfo {pages}
		{61} (\bibinfo {year} {2020})}\BibitemShut {NoStop}%
	\bibitem [{\citenamefont {Hamilton}\ \emph {et~al.}(2015)\citenamefont
		{Hamilton}, \citenamefont {Jaffe}, \citenamefont {Haslinger}, \citenamefont
		{Simmons}, \citenamefont {Müller},\ and\ \citenamefont
		{Khoury}}]{Hamilton_2015}%
	\BibitemOpen
	\bibfield  {author} {\bibinfo {author} {\bibfnamefont {P.}~\bibnamefont
			{Hamilton}}, \bibinfo {author} {\bibfnamefont {M.}~\bibnamefont {Jaffe}},
		\bibinfo {author} {\bibfnamefont {P.}~\bibnamefont {Haslinger}}, \bibinfo
		{author} {\bibfnamefont {Q.}~\bibnamefont {Simmons}}, \bibinfo {author}
		{\bibfnamefont {H.}~\bibnamefont {Müller}},\ and\ \bibinfo {author}
		{\bibfnamefont {J.}~\bibnamefont {Khoury}},\ }\bibfield  {title} {\bibinfo
		{title} {Atom-interferometry constraints on dark energy},\ }\href
	{https://doi.org/10.1126/science.aaa8883} {\bibfield  {journal} {\bibinfo
			{journal} {Science}\ }\textbf {\bibinfo {volume} {349}},\ \bibinfo {pages}
		{849} (\bibinfo {year} {2015})}\BibitemShut {NoStop}%
	\bibitem [{\citenamefont {Asenbaum}\ \emph {et~al.}(2020)\citenamefont
		{Asenbaum}, \citenamefont {Overstreet}, \citenamefont {Kim}, \citenamefont
		{Curti},\ and\ \citenamefont {Kasevich}}]{Asenbaum_2020}%
	\BibitemOpen
	\bibfield  {author} {\bibinfo {author} {\bibfnamefont {P.}~\bibnamefont
			{Asenbaum}}, \bibinfo {author} {\bibfnamefont {C.}~\bibnamefont
			{Overstreet}}, \bibinfo {author} {\bibfnamefont {M.}~\bibnamefont {Kim}},
		\bibinfo {author} {\bibfnamefont {J.}~\bibnamefont {Curti}},\ and\ \bibinfo
		{author} {\bibfnamefont {M.~A.}\ \bibnamefont {Kasevich}},\ }\bibfield
	{title} {\bibinfo {title} {Atom-interferometric test of the equivalence
			principle at the <mml:math xmlns:mml="http://www.w3.org/1998/math/mathml"},\
	}\href {https://doi.org/10.1103/PhysRevLett.125.191101} {\bibfield  {journal}
		{\bibinfo  {journal} {Physical Review Letters}\ }\textbf {\bibinfo {volume}
			{125}},\ \bibinfo {pages} {191101} (\bibinfo {year} {2020})}\BibitemShut
	{NoStop}%
	\bibitem [{\citenamefont {Fixler}\ \emph {et~al.}(2007)\citenamefont {Fixler},
		\citenamefont {Foster}, \citenamefont {McGuirk},\ and\ \citenamefont
		{Kasevich}}]{Fixler:2007is}%
	\BibitemOpen
	\bibfield  {author} {\bibinfo {author} {\bibfnamefont {J.~B.}\ \bibnamefont
			{Fixler}}, \bibinfo {author} {\bibfnamefont {G.~T.}\ \bibnamefont {Foster}},
		\bibinfo {author} {\bibfnamefont {J.~M.}\ \bibnamefont {McGuirk}},\ and\
		\bibinfo {author} {\bibfnamefont {M.~A.}\ \bibnamefont {Kasevich}},\
	}\bibfield  {title} {\bibinfo {title} {Atom interferometer measurement of the
			newtonian constant of gravity},\ }\href
	{https://doi.org/10.1126/science.1135459} {\bibfield  {journal} {\bibinfo
			{journal} {Science}\ }\textbf {\bibinfo {volume} {315}},\ \bibinfo {pages}
		{74} (\bibinfo {year} {2007})}\BibitemShut {NoStop}%
	\bibitem [{\citenamefont {Rosi}\ \emph {et~al.}(2014)\citenamefont {Rosi},
		\citenamefont {Sorrentino}, \citenamefont {Cacciapuoti}, \citenamefont
		{Prevedelli},\ and\ \citenamefont {Tino}}]{Rosi_2014}%
	\BibitemOpen
	\bibfield  {author} {\bibinfo {author} {\bibfnamefont {G.}~\bibnamefont
			{Rosi}}, \bibinfo {author} {\bibfnamefont {F.}~\bibnamefont {Sorrentino}},
		\bibinfo {author} {\bibfnamefont {L.}~\bibnamefont {Cacciapuoti}}, \bibinfo
		{author} {\bibfnamefont {M.}~\bibnamefont {Prevedelli}},\ and\ \bibinfo
		{author} {\bibfnamefont {G.~M.}\ \bibnamefont {Tino}},\ }\bibfield  {title}
	{\bibinfo {title} {Precision measurement of the newtonian gravitational
			constant using cold atoms},\ }\href {https://doi.org/10.1038/nature13433}
	{\bibfield  {journal} {\bibinfo  {journal} {Nature}\ }\textbf {\bibinfo
			{volume} {510}},\ \bibinfo {pages} {518} (\bibinfo {year}
		{2014})}\BibitemShut {NoStop}%
	\bibitem [{\citenamefont {Vessot}\ \emph {et~al.}(1980)\citenamefont {Vessot},
		\citenamefont {Levine}, \citenamefont {Mattison}, \citenamefont {Blomberg},
		\citenamefont {Hoffman}, \citenamefont {Nystrom}, \citenamefont {Farrel},
		\citenamefont {Decher}, \citenamefont {Eby}, \citenamefont {Baugher},
		\citenamefont {Watts}, \citenamefont {Teuber},\ and\ \citenamefont
		{Wills}}]{Vessot_1980}%
	\BibitemOpen
	\bibfield  {author} {\bibinfo {author} {\bibfnamefont {R.~F.~C.}\
			\bibnamefont {Vessot}}, \bibinfo {author} {\bibfnamefont {M.~W.}\
			\bibnamefont {Levine}}, \bibinfo {author} {\bibfnamefont {E.~M.}\
			\bibnamefont {Mattison}}, \bibinfo {author} {\bibfnamefont {E.~L.}\
			\bibnamefont {Blomberg}}, \bibinfo {author} {\bibfnamefont {T.~E.}\
			\bibnamefont {Hoffman}}, \bibinfo {author} {\bibfnamefont {G.~U.}\
			\bibnamefont {Nystrom}}, \bibinfo {author} {\bibfnamefont {B.~F.}\
			\bibnamefont {Farrel}}, \bibinfo {author} {\bibfnamefont {R.}~\bibnamefont
			{Decher}}, \bibinfo {author} {\bibfnamefont {P.~B.}\ \bibnamefont {Eby}},
		\bibinfo {author} {\bibfnamefont {C.~R.}\ \bibnamefont {Baugher}}, \bibinfo
		{author} {\bibfnamefont {J.~W.}\ \bibnamefont {Watts}}, \bibinfo {author}
		{\bibfnamefont {D.~L.}\ \bibnamefont {Teuber}},\ and\ \bibinfo {author}
		{\bibfnamefont {F.~D.}\ \bibnamefont {Wills}},\ }\bibfield  {title} {\bibinfo
		{title} {Test of relativistic gravitation with a space-borne hydrogen
			maser},\ }\href {https://doi.org/10.1103/PhysRevLett.45.2081} {\bibfield
		{journal} {\bibinfo  {journal} {Physical Review Letters}\ }\textbf {\bibinfo
			{volume} {45}},\ \bibinfo {pages} {2081} (\bibinfo {year}
		{1980})}\BibitemShut {NoStop}%
	\bibitem [{\citenamefont {Takamoto}\ \emph {et~al.}(2020)\citenamefont
		{Takamoto}, \citenamefont {Ushijima}, \citenamefont {Ohmae}, \citenamefont
		{Yahagi}, \citenamefont {Kokado}, \citenamefont {Shinkai},\ and\
		\citenamefont {Katori}}]{Takamoto_2020}%
	\BibitemOpen
	\bibfield  {author} {\bibinfo {author} {\bibfnamefont {M.}~\bibnamefont
			{Takamoto}}, \bibinfo {author} {\bibfnamefont {I.}~\bibnamefont {Ushijima}},
		\bibinfo {author} {\bibfnamefont {N.}~\bibnamefont {Ohmae}}, \bibinfo
		{author} {\bibfnamefont {T.}~\bibnamefont {Yahagi}}, \bibinfo {author}
		{\bibfnamefont {K.}~\bibnamefont {Kokado}}, \bibinfo {author} {\bibfnamefont
			{H.}~\bibnamefont {Shinkai}},\ and\ \bibinfo {author} {\bibfnamefont
			{H.}~\bibnamefont {Katori}},\ }\bibfield  {title} {\bibinfo {title} {Test of
			general relativity by a pair of transportable optical lattice clocks},\
	}\href {https://doi.org/10.1038/s41566-020-0619-8} {\bibfield  {journal}
		{\bibinfo  {journal} {Nature Photonics}\ }\textbf {\bibinfo {volume} {14}},\
		\bibinfo {pages} {411} (\bibinfo {year} {2020})}\BibitemShut {NoStop}%
	\bibitem [{\citenamefont {Chou}\ \emph {et~al.}(2010)\citenamefont {Chou},
		\citenamefont {Hume}, \citenamefont {Rosenband},\ and\ \citenamefont
		{Wineland}}]{Chou_2010}%
	\BibitemOpen
	\bibfield  {author} {\bibinfo {author} {\bibfnamefont {C.~W.}\ \bibnamefont
			{Chou}}, \bibinfo {author} {\bibfnamefont {D.~B.}\ \bibnamefont {Hume}},
		\bibinfo {author} {\bibfnamefont {T.}~\bibnamefont {Rosenband}},\ and\
		\bibinfo {author} {\bibfnamefont {D.~J.}\ \bibnamefont {Wineland}},\
	}\bibfield  {title} {\bibinfo {title} {Optical clocks and relativity},\
	}\href {https://doi.org/10.1126/science.1192720} {\bibfield  {journal}
		{\bibinfo  {journal} {Science}\ }\textbf {\bibinfo {volume} {329}},\ \bibinfo
		{pages} {1630} (\bibinfo {year} {2010})}\BibitemShut {NoStop}%
	\bibitem [{\citenamefont {Bothwell}\ \emph {et~al.}(2022)\citenamefont
		{Bothwell}, \citenamefont {Kennedy}, \citenamefont {Aeppli}, \citenamefont
		{Kedar}, \citenamefont {Robinson}, \citenamefont {Oelker}, \citenamefont
		{Staron},\ and\ \citenamefont {Ye}}]{Bothwell_2022}%
	\BibitemOpen
	\bibfield  {author} {\bibinfo {author} {\bibfnamefont {T.}~\bibnamefont
			{Bothwell}}, \bibinfo {author} {\bibfnamefont {C.~J.}\ \bibnamefont
			{Kennedy}}, \bibinfo {author} {\bibfnamefont {A.}~\bibnamefont {Aeppli}},
		\bibinfo {author} {\bibfnamefont {D.}~\bibnamefont {Kedar}}, \bibinfo
		{author} {\bibfnamefont {J.~M.}\ \bibnamefont {Robinson}}, \bibinfo {author}
		{\bibfnamefont {E.}~\bibnamefont {Oelker}}, \bibinfo {author} {\bibfnamefont
			{A.}~\bibnamefont {Staron}},\ and\ \bibinfo {author} {\bibfnamefont
			{J.}~\bibnamefont {Ye}},\ }\bibfield  {title} {\bibinfo {title} {Resolving
			the gravitational redshift across a millimetre-scale atomic sample},\ }\href
	{https://doi.org/10.1038/s41586-021-04349-7} {\bibfield  {journal} {\bibinfo
			{journal} {Nature}\ }\textbf {\bibinfo {volume} {602}},\ \bibinfo {pages}
		{420} (\bibinfo {year} {2022})}\BibitemShut {NoStop}%
	\bibitem [{\citenamefont {Herrmann}\ \emph {et~al.}(2018)\citenamefont
		{Herrmann}, \citenamefont {Finke}, \citenamefont {Lülf}, \citenamefont
		{Kichakova}, \citenamefont {Puetzfeld}, \citenamefont {Knickmann},
		\citenamefont {List}, \citenamefont {Rievers}, \citenamefont {Giorgi},
		\citenamefont {Günther}, \citenamefont {Dittus}, \citenamefont
		{Prieto-Cerdeira}, \citenamefont {Dilssner}, \citenamefont {Gonzalez},
		\citenamefont {Schönemann}, \citenamefont {Ventura-Traveset},\ and\
		\citenamefont {Lämmerzahl}}]{Herrmann_2018}%
	\BibitemOpen
	\bibfield  {author} {\bibinfo {author} {\bibfnamefont {S.}~\bibnamefont
			{Herrmann}}, \bibinfo {author} {\bibfnamefont {F.}~\bibnamefont {Finke}},
		\bibinfo {author} {\bibfnamefont {M.}~\bibnamefont {Lülf}}, \bibinfo
		{author} {\bibfnamefont {O.}~\bibnamefont {Kichakova}}, \bibinfo {author}
		{\bibfnamefont {D.}~\bibnamefont {Puetzfeld}}, \bibinfo {author}
		{\bibfnamefont {D.}~\bibnamefont {Knickmann}}, \bibinfo {author}
		{\bibfnamefont {M.}~\bibnamefont {List}}, \bibinfo {author} {\bibfnamefont
			{B.}~\bibnamefont {Rievers}}, \bibinfo {author} {\bibfnamefont
			{G.}~\bibnamefont {Giorgi}}, \bibinfo {author} {\bibfnamefont
			{C.}~\bibnamefont {Günther}}, \bibinfo {author} {\bibfnamefont
			{H.}~\bibnamefont {Dittus}}, \bibinfo {author} {\bibfnamefont
			{R.}~\bibnamefont {Prieto-Cerdeira}}, \bibinfo {author} {\bibfnamefont
			{F.}~\bibnamefont {Dilssner}}, \bibinfo {author} {\bibfnamefont
			{F.}~\bibnamefont {Gonzalez}}, \bibinfo {author} {\bibfnamefont
			{E.}~\bibnamefont {Schönemann}}, \bibinfo {author} {\bibfnamefont
			{J.}~\bibnamefont {Ventura-Traveset}},\ and\ \bibinfo {author} {\bibfnamefont
			{C.}~\bibnamefont {Lämmerzahl}},\ }\bibfield  {title} {\bibinfo {title}
		{Test of the gravitational redshift withgalileosatellites in an eccentric
			orbit},\ }\href {https://doi.org/10.1103/PhysRevLett.121.231102} {\bibfield
		{journal} {\bibinfo  {journal} {Physical Review Letters}\ }\textbf {\bibinfo
			{volume} {121}},\ \bibinfo {pages} {231102} (\bibinfo {year}
		{2018})}\BibitemShut {NoStop}%
	\bibitem [{\citenamefont {Delva}\ \emph {et~al.}(2018)\citenamefont {Delva},
		\citenamefont {Puchades}, \citenamefont {Schönemann}, \citenamefont
		{Dilssner}, \citenamefont {Courde}, \citenamefont {Bertone}, \citenamefont
		{Gonzalez}, \citenamefont {Hees}, \citenamefont {Le~Poncin-Lafitte},
		\citenamefont {Meynadier}, \citenamefont {Prieto-Cerdeira}, \citenamefont
		{Sohet}, \citenamefont {Ventura-Traveset},\ and\ \citenamefont
		{Wolf}}]{Delva_2018}%
	\BibitemOpen
	\bibfield  {author} {\bibinfo {author} {\bibfnamefont {P.}~\bibnamefont
			{Delva}}, \bibinfo {author} {\bibfnamefont {N.}~\bibnamefont {Puchades}},
		\bibinfo {author} {\bibfnamefont {E.}~\bibnamefont {Schönemann}}, \bibinfo
		{author} {\bibfnamefont {F.}~\bibnamefont {Dilssner}}, \bibinfo {author}
		{\bibfnamefont {C.}~\bibnamefont {Courde}}, \bibinfo {author} {\bibfnamefont
			{S.}~\bibnamefont {Bertone}}, \bibinfo {author} {\bibfnamefont
			{F.}~\bibnamefont {Gonzalez}}, \bibinfo {author} {\bibfnamefont
			{A.}~\bibnamefont {Hees}}, \bibinfo {author} {\bibfnamefont {C.}~\bibnamefont
			{Le~Poncin-Lafitte}}, \bibinfo {author} {\bibfnamefont {F.}~\bibnamefont
			{Meynadier}}, \bibinfo {author} {\bibfnamefont {R.}~\bibnamefont
			{Prieto-Cerdeira}}, \bibinfo {author} {\bibfnamefont {B.}~\bibnamefont
			{Sohet}}, \bibinfo {author} {\bibfnamefont {J.}~\bibnamefont
			{Ventura-Traveset}},\ and\ \bibinfo {author} {\bibfnamefont {P.}~\bibnamefont
			{Wolf}},\ }\bibfield  {title} {\bibinfo {title} {Gravitational redshift test
			using eccentric galileo satellites},\ }\href
	{https://doi.org/10.1103/PhysRevLett.121.231101} {\bibfield  {journal}
		{\bibinfo  {journal} {Physical Review Letters}\ }\textbf {\bibinfo {volume}
			{121}},\ \bibinfo {pages} {231101} (\bibinfo {year} {2018})}\BibitemShut
	{NoStop}%
	\bibitem [{\citenamefont {Zych}\ \emph {et~al.}()\citenamefont {Zych},
		\citenamefont {Pikovski}, \citenamefont {Costa},\ and\ \citenamefont
		{Brukner}}]{Zych_2016}%
	\BibitemOpen
	\bibfield  {author} {\bibinfo {author} {\bibfnamefont {M.}~\bibnamefont
			{Zych}}, \bibinfo {author} {\bibfnamefont {I.}~\bibnamefont {Pikovski}},
		\bibinfo {author} {\bibfnamefont {F.}~\bibnamefont {Costa}},\ and\ \bibinfo
		{author} {\bibfnamefont {C.}~\bibnamefont {Brukner}},\ }\bibfield  {title}
	{\bibinfo {title} {General relativistic effects in quantum interference of
			“clocks”},\ }\href {https://doi.org/10.1088/1742-6596/723/1/012044} {\
		\textbf {\bibinfo {volume} {723}},\ \bibinfo {pages} {012044}}\BibitemShut
	{NoStop}%
	\bibitem [{\citenamefont {Greenberger}(1970{\natexlab{a}})}]{Greenberger_1970}%
	\BibitemOpen
	\bibfield  {author} {\bibinfo {author} {\bibfnamefont {D.~M.}\ \bibnamefont
			{Greenberger}},\ }\bibfield  {title} {\bibinfo {title} {Theory of particles
			with variable mass. i. formalism},\ }\href
	{https://doi.org/10.1063/1.1665400} {\bibfield  {journal} {\bibinfo
			{journal} {Journal of Mathematical Physics}\ }\textbf {\bibinfo {volume}
			{11}},\ \bibinfo {pages} {2329} (\bibinfo {year}
		{1970}{\natexlab{a}})}\BibitemShut {NoStop}%
	\bibitem [{\citenamefont {Greenberger}(1970{\natexlab{b}})}]{Greenberger1970}%
	\BibitemOpen
	\bibfield  {author} {\bibinfo {author} {\bibfnamefont {D.~M.}\ \bibnamefont
			{Greenberger}},\ }\bibfield  {title} {\bibinfo {title} {Theory of particles
			with variable mass. ii. some physical consequences},\ }\href
	{https://doi.org/10.1063/1.1665401} {\bibfield  {journal} {\bibinfo
			{journal} {Journal of Mathematical Physics}\ }\textbf {\bibinfo {volume}
			{11}},\ \bibinfo {pages} {2341} (\bibinfo {year}
		{1970}{\natexlab{b}})}\BibitemShut {NoStop}%
	\bibitem [{\citenamefont {Zych}\ \emph {et~al.}(2011)\citenamefont {Zych},
		\citenamefont {Costa}, \citenamefont {Pikovski},\ and\ \citenamefont
		{Brukner}}]{Zych_2011}%
	\BibitemOpen
	\bibfield  {author} {\bibinfo {author} {\bibfnamefont {M.}~\bibnamefont
			{Zych}}, \bibinfo {author} {\bibfnamefont {F.}~\bibnamefont {Costa}},
		\bibinfo {author} {\bibfnamefont {I.}~\bibnamefont {Pikovski}},\ and\
		\bibinfo {author} {\bibfnamefont {C.}~\bibnamefont {Brukner}},\ }\bibfield
	{title} {\bibinfo {title} {Quantum interferometric visibility as a witness of
			general relativistic proper time},\ }\bibfield  {journal} {\bibinfo
		{journal} {Nature Communications}\ }\textbf {\bibinfo {volume} {2}},\ \href
	{https://doi.org/10.1038/ncomms1498} {10.1038/ncomms1498} (\bibinfo {year}
	{2011})\BibitemShut {NoStop}%
	\bibitem [{\citenamefont {Pikovski}\ \emph {et~al.}(2015)\citenamefont
		{Pikovski}, \citenamefont {Zych}, \citenamefont {Costa},\ and\ \citenamefont
		{Brukner}}]{Pikovski_2015}%
	\BibitemOpen
	\bibfield  {author} {\bibinfo {author} {\bibfnamefont {I.}~\bibnamefont
			{Pikovski}}, \bibinfo {author} {\bibfnamefont {M.}~\bibnamefont {Zych}},
		\bibinfo {author} {\bibfnamefont {F.}~\bibnamefont {Costa}},\ and\ \bibinfo
		{author} {\bibfnamefont {C.}~\bibnamefont {Brukner}},\ }\bibfield  {title}
	{\bibinfo {title} {Universal decoherence due to gravitational
			time dilation},\ }\href {https://doi.org/10.1038/nphys3366} {\bibfield
		{journal} {\bibinfo  {journal} {Nature Physics}\ }\textbf {\bibinfo {volume}
			{11}},\ \bibinfo {pages} {668} (\bibinfo {year} {2015})}\BibitemShut
	{NoStop}%
	\bibitem [{\citenamefont {Bassi}\ \emph {et~al.}(2017)\citenamefont {Bassi},
		\citenamefont {Großardt},\ and\ \citenamefont {Ulbricht}}]{Bassi_2017}%
	\BibitemOpen
	\bibfield  {author} {\bibinfo {author} {\bibfnamefont {A.}~\bibnamefont
			{Bassi}}, \bibinfo {author} {\bibfnamefont {A.}~\bibnamefont {Großardt}},\
		and\ \bibinfo {author} {\bibfnamefont {H.}~\bibnamefont {Ulbricht}},\
	}\bibfield  {title} {\bibinfo {title} {Gravitational decoherence},\ }\href
	{https://doi.org/10.1088/1361-6382/aa864f} {\bibfield  {journal} {\bibinfo
			{journal} {Classical and Quantum Gravity}\ }\textbf {\bibinfo {volume}
			{34}},\ \bibinfo {pages} {193002} (\bibinfo {year} {2017})}\BibitemShut
	{NoStop}%
	\bibitem [{\citenamefont {Schleich}\ \emph {et~al.}(2013)\citenamefont
		{Schleich}, \citenamefont {Greenberger},\ and\ \citenamefont
		{Rasel}}]{Schleich_2013}%
	\BibitemOpen
	\bibfield  {author} {\bibinfo {author} {\bibfnamefont {W.~P.}\ \bibnamefont
			{Schleich}}, \bibinfo {author} {\bibfnamefont {D.~M.}\ \bibnamefont
			{Greenberger}},\ and\ \bibinfo {author} {\bibfnamefont {E.~M.}\ \bibnamefont
			{Rasel}},\ }\bibfield  {title} {\bibinfo {title} {A representation-free
			description of the kasevich–chu interferometer: a resolution of the
			redshift controversy},\ }\href
	{https://doi.org/10.1088/1367-2630/15/1/013007} {\bibfield  {journal}
		{\bibinfo  {journal} {New Journal of Physics}\ }\textbf {\bibinfo {volume}
			{15}},\ \bibinfo {pages} {013007} (\bibinfo {year} {2013})}\BibitemShut
	{NoStop}%
	\bibitem [{\citenamefont {Müller}\ \emph {et~al.}(2010)\citenamefont
		{Müller}, \citenamefont {Peters},\ and\ \citenamefont {Chu}}]{M_ller_2010}%
	\BibitemOpen
	\bibfield  {author} {\bibinfo {author} {\bibfnamefont {H.}~\bibnamefont
			{Müller}}, \bibinfo {author} {\bibfnamefont {A.}~\bibnamefont {Peters}},\
		and\ \bibinfo {author} {\bibfnamefont {S.}~\bibnamefont {Chu}},\ }\bibfield
	{title} {\bibinfo {title} {A precision measurement of the gravitational
			redshift by the interference of matter waves},\ }\href
	{https://doi.org/10.1038/nature08776} {\bibfield  {journal} {\bibinfo
			{journal} {Nature}\ }\textbf {\bibinfo {volume} {463}},\ \bibinfo {pages}
		{926} (\bibinfo {year} {2010})}\BibitemShut {NoStop}%
	\bibitem [{\citenamefont {Wolf}\ \emph {et~al.}(2011)\citenamefont {Wolf},
		\citenamefont {Blanchet}, \citenamefont {Bordé}, \citenamefont {Reynaud},
		\citenamefont {Salomon},\ and\ \citenamefont {Cohen-Tannoudji}}]{Wolf_2011}%
	\BibitemOpen
	\bibfield  {author} {\bibinfo {author} {\bibfnamefont {P.}~\bibnamefont
			{Wolf}}, \bibinfo {author} {\bibfnamefont {L.}~\bibnamefont {Blanchet}},
		\bibinfo {author} {\bibfnamefont {C.~J.}\ \bibnamefont {Bordé}}, \bibinfo
		{author} {\bibfnamefont {S.}~\bibnamefont {Reynaud}}, \bibinfo {author}
		{\bibfnamefont {C.}~\bibnamefont {Salomon}},\ and\ \bibinfo {author}
		{\bibfnamefont {C.}~\bibnamefont {Cohen-Tannoudji}},\ }\bibfield  {title}
	{\bibinfo {title} {Does an atom interferometer test the gravitational
			redshift at the compton frequency?},\ }\href
	{https://doi.org/10.1088/0264-9381/28/14/145017} {\bibfield  {journal}
		{\bibinfo  {journal} {Classical and Quantum Gravity}\ }\textbf {\bibinfo
			{volume} {28}},\ \bibinfo {pages} {145017} (\bibinfo {year}
		{2011})}\BibitemShut {NoStop}%
	\bibitem [{\citenamefont {Rosi}\ \emph {et~al.}(2017)\citenamefont {Rosi},
		\citenamefont {D’Amico}, \citenamefont {Cacciapuoti}, \citenamefont
		{Sorrentino}, \citenamefont {Prevedelli}, \citenamefont {Zych}, \citenamefont
		{Brukner},\ and\ \citenamefont {Tino}}]{Rosi_2017}%
	\BibitemOpen
	\bibfield  {author} {\bibinfo {author} {\bibfnamefont {G.}~\bibnamefont
			{Rosi}}, \bibinfo {author} {\bibfnamefont {G.}~\bibnamefont {D’Amico}},
		\bibinfo {author} {\bibfnamefont {L.}~\bibnamefont {Cacciapuoti}}, \bibinfo
		{author} {\bibfnamefont {F.}~\bibnamefont {Sorrentino}}, \bibinfo {author}
		{\bibfnamefont {M.}~\bibnamefont {Prevedelli}}, \bibinfo {author}
		{\bibfnamefont {M.}~\bibnamefont {Zych}}, \bibinfo {author} {\bibfnamefont
			{C.}~\bibnamefont {Brukner}},\ and\ \bibinfo {author} {\bibfnamefont {G.~M.}\
			\bibnamefont {Tino}},\ }\bibfield  {title} {\bibinfo {title} {Quantum test of
			the equivalence principle for atoms in coherent superposition of internal
			energy states},\ }\bibfield  {journal} {\bibinfo  {journal} {Nature
			Communications}\ }\textbf {\bibinfo {volume} {8}},\ \href
	{https://doi.org/10.1038/ncomms15529} {10.1038/ncomms15529} (\bibinfo {year}
	{2017})\BibitemShut {NoStop}%
	\bibitem [{\citenamefont {Ludlow}\ \emph {et~al.}(2015)\citenamefont {Ludlow},
		\citenamefont {Boyd}, \citenamefont {Ye}, \citenamefont {Peik},\ and\
		\citenamefont {Schmidt}}]{Ludlow_2015}%
	\BibitemOpen
	\bibfield  {author} {\bibinfo {author} {\bibfnamefont {A.~D.}\ \bibnamefont
			{Ludlow}}, \bibinfo {author} {\bibfnamefont {M.~M.}\ \bibnamefont {Boyd}},
		\bibinfo {author} {\bibfnamefont {J.}~\bibnamefont {Ye}}, \bibinfo {author}
		{\bibfnamefont {E.}~\bibnamefont {Peik}},\ and\ \bibinfo {author}
		{\bibfnamefont {P.}~\bibnamefont {Schmidt}},\ }\bibfield  {title} {\bibinfo
		{title} {Optical atomic clocks},\ }\href
	{https://doi.org/10.1103/RevModPhys.87.637} {\bibfield  {journal} {\bibinfo
			{journal} {Reviews of Modern Physics}\ }\textbf {\bibinfo {volume} {87}},\
		\bibinfo {pages} {637} (\bibinfo {year} {2015})}\BibitemShut {NoStop}%
	\bibitem [{\citenamefont {Metcalf}\ and\ \citenamefont {van~der
			Straten}(1999)}]{Metcalf_1999}%
	\BibitemOpen
	\bibfield  {author} {\bibinfo {author} {\bibfnamefont {H.~J.}\ \bibnamefont
			{Metcalf}}\ and\ \bibinfo {author} {\bibfnamefont {P.}~\bibnamefont {van~der
				Straten}},\ }\href {https://doi.org/10.1007/978-1-4612-1470-0} {\emph
		{\bibinfo {title} {Laser Cooling and Trapping}}}\ (\bibinfo  {publisher}
	{Springer New York},\ \bibinfo {year} {1999})\BibitemShut {NoStop}%
	\bibitem [{\citenamefont {Young}\ \emph {et~al.}(2020)\citenamefont {Young},
		\citenamefont {Eckner}, \citenamefont {Milner}, \citenamefont {Kedar},
		\citenamefont {Norcia}, \citenamefont {Oelker}, \citenamefont {Schine},
		\citenamefont {Ye},\ and\ \citenamefont {Kaufman}}]{Young_2020}%
	\BibitemOpen
	\bibfield  {author} {\bibinfo {author} {\bibfnamefont {A.~W.}\ \bibnamefont
			{Young}}, \bibinfo {author} {\bibfnamefont {W.~J.}\ \bibnamefont {Eckner}},
		\bibinfo {author} {\bibfnamefont {W.~R.}\ \bibnamefont {Milner}}, \bibinfo
		{author} {\bibfnamefont {D.}~\bibnamefont {Kedar}}, \bibinfo {author}
		{\bibfnamefont {M.~A.}\ \bibnamefont {Norcia}}, \bibinfo {author}
		{\bibfnamefont {E.}~\bibnamefont {Oelker}}, \bibinfo {author} {\bibfnamefont
			{N.}~\bibnamefont {Schine}}, \bibinfo {author} {\bibfnamefont
			{J.}~\bibnamefont {Ye}},\ and\ \bibinfo {author} {\bibfnamefont {A.~M.}\
			\bibnamefont {Kaufman}},\ }\bibfield  {title} {\bibinfo {title}
		{Half-minute-scale atomic coherence and high relative stability in a tweezer
			clock},\ }\href {https://doi.org/10.1038/s41586-020-3009-y} {\bibfield
		{journal} {\bibinfo  {journal} {Nature}\ }\textbf {\bibinfo {volume} {588}},\
		\bibinfo {pages} {408} (\bibinfo {year} {2020})}\BibitemShut {NoStop}%
	\bibitem [{\citenamefont {Hu}\ \emph {et~al.}(2017)\citenamefont {Hu},
		\citenamefont {Poli}, \citenamefont {Salvi},\ and\ \citenamefont
		{Tino}}]{Hu_2017}%
	\BibitemOpen
	\bibfield  {author} {\bibinfo {author} {\bibfnamefont {L.}~\bibnamefont
			{Hu}}, \bibinfo {author} {\bibfnamefont {N.}~\bibnamefont {Poli}}, \bibinfo
		{author} {\bibfnamefont {L.}~\bibnamefont {Salvi}},\ and\ \bibinfo {author}
		{\bibfnamefont {G.~M.}\ \bibnamefont {Tino}},\ }\bibfield  {title} {\bibinfo
		{title} {Atom interferometry with the sr optical clock transition},\ }\href
	{https://doi.org/10.1103/PhysRevLett.119.263601} {\bibfield  {journal}
		{\bibinfo  {journal} {Physical Review Letters}\ }\textbf {\bibinfo {volume}
			{119}},\ \bibinfo {pages} {263601} (\bibinfo {year} {2017})}\BibitemShut
	{NoStop}%
	\bibitem [{\citenamefont {Hu}\ \emph {et~al.}(2019)\citenamefont {Hu},
		\citenamefont {Wang}, \citenamefont {Salvi}, \citenamefont {Tinsley},
		\citenamefont {Tino},\ and\ \citenamefont {Poli}}]{Hu_2019}%
	\BibitemOpen
	\bibfield  {author} {\bibinfo {author} {\bibfnamefont {L.}~\bibnamefont
			{Hu}}, \bibinfo {author} {\bibfnamefont {E.}~\bibnamefont {Wang}}, \bibinfo
		{author} {\bibfnamefont {L.}~\bibnamefont {Salvi}}, \bibinfo {author}
		{\bibfnamefont {J.~N.}\ \bibnamefont {Tinsley}}, \bibinfo {author}
		{\bibfnamefont {G.~M.}\ \bibnamefont {Tino}},\ and\ \bibinfo {author}
		{\bibfnamefont {N.}~\bibnamefont {Poli}},\ }\bibfield  {title} {\bibinfo
		{title} {Sr atom interferometry with the optical clock transition as a
			gravimeter and a gravity gradiometer},\ }\href
	{https://doi.org/10.1088/1361-6382/ab4d18} {\bibfield  {journal} {\bibinfo
			{journal} {Classical and Quantum Gravity}\ }\textbf {\bibinfo {volume}
			{37}},\ \bibinfo {pages} {014001} (\bibinfo {year} {2019})}\BibitemShut
	{NoStop}%
	\bibitem [{\citenamefont {Rudolph}\ \emph {et~al.}(2020)\citenamefont
		{Rudolph}, \citenamefont {Wilkason}, \citenamefont {Nantel}, \citenamefont
		{Swan}, \citenamefont {Holland}, \citenamefont {Jiang}, \citenamefont
		{Garber}, \citenamefont {Carman},\ and\ \citenamefont
		{Hogan}}]{Rudolph_2020}%
	\BibitemOpen
	\bibfield  {author} {\bibinfo {author} {\bibfnamefont {J.}~\bibnamefont
			{Rudolph}}, \bibinfo {author} {\bibfnamefont {T.}~\bibnamefont {Wilkason}},
		\bibinfo {author} {\bibfnamefont {M.}~\bibnamefont {Nantel}}, \bibinfo
		{author} {\bibfnamefont {H.}~\bibnamefont {Swan}}, \bibinfo {author}
		{\bibfnamefont {C.~M.}\ \bibnamefont {Holland}}, \bibinfo {author}
		{\bibfnamefont {Y.}~\bibnamefont {Jiang}}, \bibinfo {author} {\bibfnamefont
			{B.~E.}\ \bibnamefont {Garber}}, \bibinfo {author} {\bibfnamefont {S.~P.}\
			\bibnamefont {Carman}},\ and\ \bibinfo {author} {\bibfnamefont {J.~M.}\
			\bibnamefont {Hogan}},\ }\bibfield  {title} {\bibinfo {title} {Large momentum
			transfer clock atom interferometry on the 689 nm intercombination line of
			strontium},\ }\href {https://doi.org/10.1103/PhysRevLett.124.083604}
	{\bibfield  {journal} {\bibinfo  {journal} {Physical Review Letters}\
		}\textbf {\bibinfo {volume} {124}},\ \bibinfo {pages} {083604} (\bibinfo
		{year} {2020})}\BibitemShut {NoStop}%
	\bibitem [{\citenamefont {Roura}(2020)}]{Roura_2020}%
	\BibitemOpen
	\bibfield  {author} {\bibinfo {author} {\bibfnamefont {A.}~\bibnamefont
			{Roura}},\ }\bibfield  {title} {\bibinfo {title} {Gravitational redshift in
			quantum-clock interferometry},\ }\href
	{https://doi.org/10.1103/PhysRevX.10.021014} {\bibfield  {journal} {\bibinfo
			{journal} {Physical Review X}\ }\textbf {\bibinfo {volume} {10}},\ \bibinfo
		{pages} {021014} (\bibinfo {year} {2020})}\BibitemShut {NoStop}%
	\bibitem [{\citenamefont {Di~Pumpo}\ \emph {et~al.}(2021)\citenamefont
		{Di~Pumpo}, \citenamefont {Ufrecht}, \citenamefont {Friedrich}, \citenamefont
		{Giese}, \citenamefont {Schleich},\ and\ \citenamefont
		{Unruh}}]{Di_Pumpo_2021}%
	\BibitemOpen
	\bibfield  {author} {\bibinfo {author} {\bibfnamefont {F.}~\bibnamefont
			{Di~Pumpo}}, \bibinfo {author} {\bibfnamefont {C.}~\bibnamefont {Ufrecht}},
		\bibinfo {author} {\bibfnamefont {A.}~\bibnamefont {Friedrich}}, \bibinfo
		{author} {\bibfnamefont {E.}~\bibnamefont {Giese}}, \bibinfo {author}
		{\bibfnamefont {W.~P.}\ \bibnamefont {Schleich}},\ and\ \bibinfo {author}
		{\bibfnamefont {W.~G.}\ \bibnamefont {Unruh}},\ }\bibfield  {title} {\bibinfo
		{title} {Gravitational redshift tests with atomic clocks and atom
			interferometers},\ }\href {https://doi.org/10.1103/PRXQuantum.2.040333}
	{\bibfield  {journal} {\bibinfo  {journal} {PRX Quantum}\ }\textbf {\bibinfo
			{volume} {2}},\ \bibinfo {pages} {040333} (\bibinfo {year}
		{2021})}\BibitemShut {NoStop}%
	\bibitem [{\citenamefont {Giese}\ \emph {et~al.}(2019)\citenamefont {Giese},
		\citenamefont {Friedrich}, \citenamefont {Di~Pumpo}, \citenamefont {Roura},
		\citenamefont {Schleich}, \citenamefont {Greenberger},\ and\ \citenamefont
		{Rasel}}]{Giese_2019}%
	\BibitemOpen
	\bibfield  {author} {\bibinfo {author} {\bibfnamefont {E.}~\bibnamefont
			{Giese}}, \bibinfo {author} {\bibfnamefont {A.}~\bibnamefont {Friedrich}},
		\bibinfo {author} {\bibfnamefont {F.}~\bibnamefont {Di~Pumpo}}, \bibinfo
		{author} {\bibfnamefont {A.}~\bibnamefont {Roura}}, \bibinfo {author}
		{\bibfnamefont {W.~P.}\ \bibnamefont {Schleich}}, \bibinfo {author}
		{\bibfnamefont {D.~M.}\ \bibnamefont {Greenberger}},\ and\ \bibinfo {author}
		{\bibfnamefont {E.~M.}\ \bibnamefont {Rasel}},\ }\bibfield  {title} {\bibinfo
		{title} {Proper time in atom interferometers: Diffractive versus specular
			mirrors},\ }\href {https://doi.org/10.1103/PhysRevA.99.013627} {\bibfield
		{journal} {\bibinfo  {journal} {Physical Review A}\ }\textbf {\bibinfo
			{volume} {99}},\ \bibinfo {pages} {013627} (\bibinfo {year}
		{2019})}\BibitemShut {NoStop}%
	\bibitem [{\citenamefont {Greenberger}\ \emph {et~al.}(2012)\citenamefont
		{Greenberger}, \citenamefont {Schleich},\ and\ \citenamefont
		{Rasel}}]{Greenberger_2012}%
	\BibitemOpen
	\bibfield  {author} {\bibinfo {author} {\bibfnamefont {D.~M.}\ \bibnamefont
			{Greenberger}}, \bibinfo {author} {\bibfnamefont {W.~P.}\ \bibnamefont
			{Schleich}},\ and\ \bibinfo {author} {\bibfnamefont {E.~M.}\ \bibnamefont
			{Rasel}},\ }\bibfield  {title} {\bibinfo {title} {Relativistic effects in
			atom and neutron interferometry and the differences between them},\ }\href
	{https://doi.org/10.1103/PhysRevA.86.063622} {\bibfield  {journal} {\bibinfo
			{journal} {Physical Review A}\ }\textbf {\bibinfo {volume} {86}},\ \bibinfo
		{pages} {063622} (\bibinfo {year} {2012})}\BibitemShut {NoStop}%
	\bibitem [{\citenamefont {Loriani}\ \emph {et~al.}(2019)\citenamefont
		{Loriani}, \citenamefont {Friedrich}, \citenamefont {Ufrecht}, \citenamefont
		{Di~Pumpo}, \citenamefont {Kleinert}, \citenamefont {Abend}, \citenamefont
		{Gaaloul}, \citenamefont {Meiners}, \citenamefont {Schubert}, \citenamefont
		{Tell}, \citenamefont {Wodey}, \citenamefont {Zych}, \citenamefont {Ertmer},
		\citenamefont {Roura}, \citenamefont {Schlippert}, \citenamefont {Schleich},
		\citenamefont {Rasel},\ and\ \citenamefont {Giese}}]{Loriani_2019}%
	\BibitemOpen
	\bibfield  {author} {\bibinfo {author} {\bibfnamefont {S.}~\bibnamefont
			{Loriani}}, \bibinfo {author} {\bibfnamefont {A.}~\bibnamefont {Friedrich}},
		\bibinfo {author} {\bibfnamefont {C.}~\bibnamefont {Ufrecht}}, \bibinfo
		{author} {\bibfnamefont {F.}~\bibnamefont {Di~Pumpo}}, \bibinfo {author}
		{\bibfnamefont {S.}~\bibnamefont {Kleinert}}, \bibinfo {author}
		{\bibfnamefont {S.}~\bibnamefont {Abend}}, \bibinfo {author} {\bibfnamefont
			{N.}~\bibnamefont {Gaaloul}}, \bibinfo {author} {\bibfnamefont
			{C.}~\bibnamefont {Meiners}}, \bibinfo {author} {\bibfnamefont
			{C.}~\bibnamefont {Schubert}}, \bibinfo {author} {\bibfnamefont
			{D.}~\bibnamefont {Tell}}, \bibinfo {author} {\bibfnamefont {E.}~\bibnamefont
			{Wodey}}, \bibinfo {author} {\bibfnamefont {M.}~\bibnamefont {Zych}},
		\bibinfo {author} {\bibfnamefont {W.}~\bibnamefont {Ertmer}}, \bibinfo
		{author} {\bibfnamefont {A.}~\bibnamefont {Roura}}, \bibinfo {author}
		{\bibfnamefont {D.}~\bibnamefont {Schlippert}}, \bibinfo {author}
		{\bibfnamefont {W.~P.}\ \bibnamefont {Schleich}}, \bibinfo {author}
		{\bibfnamefont {E.~M.}\ \bibnamefont {Rasel}},\ and\ \bibinfo {author}
		{\bibfnamefont {E.}~\bibnamefont {Giese}},\ }\bibfield  {title} {\bibinfo
		{title} {Interference of clocks: A quantum twin paradox},\ }\bibfield
	{journal} {\bibinfo  {journal} {Science Advances}\ }\textbf {\bibinfo
		{volume} {5}},\ \href {https://doi.org/10.1126/sciadv.aax8966}
	{10.1126/sciadv.aax8966} (\bibinfo {year} {2019})\BibitemShut {NoStop}%
	\bibitem [{\citenamefont {Florshaim}\ \emph {et~al.}(2023)\citenamefont
		{Florshaim}, \citenamefont {Zohar}, \citenamefont {Koplovich}, \citenamefont
		{Meltzer}, \citenamefont {Weill}, \citenamefont {Nemirovsky}, \citenamefont
		{Stern},\ and\ \citenamefont {Sagi}}]{Florshaim2023}%
	\BibitemOpen
	\bibfield  {author} {\bibinfo {author} {\bibfnamefont {Y.}~\bibnamefont
			{Florshaim}}, \bibinfo {author} {\bibfnamefont {E.}~\bibnamefont {Zohar}},
		\bibinfo {author} {\bibfnamefont {D.~Z.}\ \bibnamefont {Koplovich}}, \bibinfo
		{author} {\bibfnamefont {I.}~\bibnamefont {Meltzer}}, \bibinfo {author}
		{\bibfnamefont {R.}~\bibnamefont {Weill}}, \bibinfo {author} {\bibfnamefont
			{J.}~\bibnamefont {Nemirovsky}}, \bibinfo {author} {\bibfnamefont
			{A.}~\bibnamefont {Stern}},\ and\ \bibinfo {author} {\bibfnamefont
			{Y.}~\bibnamefont {Sagi}},\ }\bibfield  {title} {\bibinfo {title} {Spatial
			adiabatic passage of ultracold atoms in optical tweezers}\ }\href
	{https://doi.org/10.48550/arXiv.2305.16228} {10.48550/arXiv.2305.16228}
	(\bibinfo {year} {2023})\BibitemShut {NoStop}%
	\bibitem [{\citenamefont {Serwane}\ \emph {et~al.}(2011)\citenamefont
		{Serwane}, \citenamefont {Zürn}, \citenamefont {Lompe}, \citenamefont
		{Ottenstein}, \citenamefont {Wenz},\ and\ \citenamefont
		{Jochim}}]{Serwane_2011}%
	\BibitemOpen
	\bibfield  {author} {\bibinfo {author} {\bibfnamefont {F.}~\bibnamefont
			{Serwane}}, \bibinfo {author} {\bibfnamefont {G.}~\bibnamefont {Zürn}},
		\bibinfo {author} {\bibfnamefont {T.}~\bibnamefont {Lompe}}, \bibinfo
		{author} {\bibfnamefont {T.~B.}\ \bibnamefont {Ottenstein}}, \bibinfo
		{author} {\bibfnamefont {A.~N.}\ \bibnamefont {Wenz}},\ and\ \bibinfo
		{author} {\bibfnamefont {S.}~\bibnamefont {Jochim}},\ }\bibfield  {title}
	{\bibinfo {title} {Deterministic preparation of a tunable few-fermion
			system},\ }\href {https://doi.org/10.1126/science.1201351} {\bibfield
		{journal} {\bibinfo  {journal} {Science}\ }\textbf {\bibinfo {volume}
			{332}},\ \bibinfo {pages} {336} (\bibinfo {year} {2011})}\BibitemShut
	{NoStop}%
	\bibitem [{\citenamefont {Tino}(2021)}]{Tino_2021}%
	\BibitemOpen
	\bibfield  {author} {\bibinfo {author} {\bibfnamefont {G.~M.}\ \bibnamefont
			{Tino}},\ }\bibfield  {title} {\bibinfo {title} {Testing gravity with cold
			atom interferometry: results and prospects},\ }\href
	{https://doi.org/10.1088/2058-9565/abd83e} {\bibfield  {journal} {\bibinfo
			{journal} {Quantum Science and Technology}\ }\textbf {\bibinfo {volume}
			{6}},\ \bibinfo {pages} {024014} (\bibinfo {year} {2021})}\BibitemShut
	{NoStop}%
	\bibitem [{\citenamefont {Marletto}\ and\ \citenamefont
		{Vedral}(2017)}]{Marletto_2017}%
	\BibitemOpen
	\bibfield  {author} {\bibinfo {author} {\bibfnamefont {C.}~\bibnamefont
			{Marletto}}\ and\ \bibinfo {author} {\bibfnamefont {V.}~\bibnamefont
			{Vedral}},\ }\bibfield  {title} {\bibinfo {title} {Gravitationally induced
			entanglement between two massive particles is sufficient evidence of quantum
			effects in gravity},\ }\href {https://doi.org/10.1103/PhysRevLett.119.240402}
	{\bibfield  {journal} {\bibinfo  {journal} {Physical Review Letters}\
		}\textbf {\bibinfo {volume} {119}},\ \bibinfo {pages} {240402} (\bibinfo
		{year} {2017})}\BibitemShut {NoStop}%
	\bibitem [{\citenamefont {Bose}\ \emph {et~al.}(2017)\citenamefont {Bose},
		\citenamefont {Mazumdar}, \citenamefont {Morley}, \citenamefont {Ulbricht},
		\citenamefont {Toroš}, \citenamefont {Paternostro}, \citenamefont {Geraci},
		\citenamefont {Barker}, \citenamefont {Kim},\ and\ \citenamefont
		{Milburn}}]{Bose_2017}%
	\BibitemOpen
	\bibfield  {author} {\bibinfo {author} {\bibfnamefont {S.}~\bibnamefont
			{Bose}}, \bibinfo {author} {\bibfnamefont {A.}~\bibnamefont {Mazumdar}},
		\bibinfo {author} {\bibfnamefont {G.~W.}\ \bibnamefont {Morley}}, \bibinfo
		{author} {\bibfnamefont {H.}~\bibnamefont {Ulbricht}}, \bibinfo {author}
		{\bibfnamefont {M.}~\bibnamefont {Toroš}}, \bibinfo {author} {\bibfnamefont
			{M.}~\bibnamefont {Paternostro}}, \bibinfo {author} {\bibfnamefont {A.~A.}\
			\bibnamefont {Geraci}}, \bibinfo {author} {\bibfnamefont {P.~F.}\
			\bibnamefont {Barker}}, \bibinfo {author} {\bibfnamefont {M.}~\bibnamefont
			{Kim}},\ and\ \bibinfo {author} {\bibfnamefont {G.}~\bibnamefont {Milburn}},\
	}\bibfield  {title} {\bibinfo {title} {Spin entanglement witness for quantum
			gravity},\ }\href {https://doi.org/10.1103/PhysRevLett.119.240401} {\bibfield
		{journal} {\bibinfo  {journal} {Physical Review Letters}\ }\textbf {\bibinfo
			{volume} {119}},\ \bibinfo {pages} {240401} (\bibinfo {year}
		{2017})}\BibitemShut {NoStop}%
	\bibitem [{\citenamefont {Sinha}\ and\ \citenamefont
		{Samuel}(2011)}]{Sinha_2011}%
	\BibitemOpen
	\bibfield  {author} {\bibinfo {author} {\bibfnamefont {S.}~\bibnamefont
			{Sinha}}\ and\ \bibinfo {author} {\bibfnamefont {J.}~\bibnamefont {Samuel}},\
	}\bibfield  {title} {\bibinfo {title} {Atom interferometry and the
			gravitational redshift},\ }\href
	{https://doi.org/10.1088/0264-9381/28/14/145018} {\bibfield  {journal}
		{\bibinfo  {journal} {Classical and Quantum Gravity}\ }\textbf {\bibinfo
			{volume} {28}},\ \bibinfo {pages} {145018} (\bibinfo {year}
		{2011})}\BibitemShut {NoStop}%
	\bibitem [{\citenamefont {Overstreet}\ \emph {et~al.}(2022)\citenamefont
		{Overstreet}, \citenamefont {Asenbaum}, \citenamefont {Curti}, \citenamefont
		{Kim},\ and\ \citenamefont {Kasevich}}]{Overstreet_2022}%
	\BibitemOpen
	\bibfield  {author} {\bibinfo {author} {\bibfnamefont {C.}~\bibnamefont
			{Overstreet}}, \bibinfo {author} {\bibfnamefont {P.}~\bibnamefont
			{Asenbaum}}, \bibinfo {author} {\bibfnamefont {J.}~\bibnamefont {Curti}},
		\bibinfo {author} {\bibfnamefont {M.}~\bibnamefont {Kim}},\ and\ \bibinfo
		{author} {\bibfnamefont {M.~A.}\ \bibnamefont {Kasevich}},\ }\bibfield
	{title} {\bibinfo {title} {Observation of a gravitational aharonov-bohm
			effect},\ }\href {https://doi.org/10.1126/science.abl7152} {\bibfield
		{journal} {\bibinfo  {journal} {Science}\ }\textbf {\bibinfo {volume}
			{375}},\ \bibinfo {pages} {226} (\bibinfo {year} {2022})}\BibitemShut
	{NoStop}%
	\bibitem [{\citenamefont {Colella}\ \emph {et~al.}(1975)\citenamefont
		{Colella}, \citenamefont {Overhauser},\ and\ \citenamefont
		{Werner}}]{Colella_1975}%
	\BibitemOpen
	\bibfield  {author} {\bibinfo {author} {\bibfnamefont {R.}~\bibnamefont
			{Colella}}, \bibinfo {author} {\bibfnamefont {A.~W.}\ \bibnamefont
			{Overhauser}},\ and\ \bibinfo {author} {\bibfnamefont {S.~A.}\ \bibnamefont
			{Werner}},\ }\bibfield  {title} {\bibinfo {title} {Observation of
			gravitationally induced quantum interference},\ }\href
	{https://doi.org/10.1103/PhysRevLett.34.1472} {\bibfield  {journal} {\bibinfo
			{journal} {Physical Review Letters}\ }\textbf {\bibinfo {volume} {34}},\
		\bibinfo {pages} {1472} (\bibinfo {year} {1975})}\BibitemShut {NoStop}%
	\bibitem [{\citenamefont {Hinkley}\ \emph {et~al.}(2013)\citenamefont
		{Hinkley}, \citenamefont {Sherman}, \citenamefont {Phillips}, \citenamefont
		{Schioppo}, \citenamefont {Lemke}, \citenamefont {Beloy}, \citenamefont
		{Pizzocaro}, \citenamefont {Oates},\ and\ \citenamefont
		{Ludlow}}]{Hinkley_2013}%
	\BibitemOpen
	\bibfield  {author} {\bibinfo {author} {\bibfnamefont {N.}~\bibnamefont
			{Hinkley}}, \bibinfo {author} {\bibfnamefont {J.~A.}\ \bibnamefont
			{Sherman}}, \bibinfo {author} {\bibfnamefont {N.~B.}\ \bibnamefont
			{Phillips}}, \bibinfo {author} {\bibfnamefont {M.}~\bibnamefont {Schioppo}},
		\bibinfo {author} {\bibfnamefont {N.~D.}\ \bibnamefont {Lemke}}, \bibinfo
		{author} {\bibfnamefont {K.}~\bibnamefont {Beloy}}, \bibinfo {author}
		{\bibfnamefont {M.}~\bibnamefont {Pizzocaro}}, \bibinfo {author}
		{\bibfnamefont {C.~W.}\ \bibnamefont {Oates}},\ and\ \bibinfo {author}
		{\bibfnamefont {A.~D.}\ \bibnamefont {Ludlow}},\ }\bibfield  {title}
	{\bibinfo {title} {An atomic clock with 10 –18 instability},\ }\href
	{https://doi.org/10.1126/science.1240420} {\bibfield  {journal} {\bibinfo
			{journal} {Science}\ }\textbf {\bibinfo {volume} {341}},\ \bibinfo {pages}
		{1215} (\bibinfo {year} {2013})}\BibitemShut {NoStop}%
	\bibitem [{\citenamefont {Katori}\ \emph {et~al.}(2015)\citenamefont {Katori},
		\citenamefont {Ovsiannikov}, \citenamefont {Marmo},\ and\ \citenamefont
		{Palchikov}}]{Katori_2015}%
	\BibitemOpen
	\bibfield  {author} {\bibinfo {author} {\bibfnamefont {H.}~\bibnamefont
			{Katori}}, \bibinfo {author} {\bibfnamefont {V.~D.}\ \bibnamefont
			{Ovsiannikov}}, \bibinfo {author} {\bibfnamefont {S.~I.}\ \bibnamefont
			{Marmo}},\ and\ \bibinfo {author} {\bibfnamefont {V.~G.}\ \bibnamefont
			{Palchikov}},\ }\bibfield  {title} {\bibinfo {title} {Strategies for reducing
			the light shift in atomic clocks},\ }\href
	{https://doi.org/10.1103/PhysRevA.91.052503} {\bibfield  {journal} {\bibinfo
			{journal} {Physical Review A}\ }\textbf {\bibinfo {volume} {91}},\ \bibinfo
		{pages} {052503} (\bibinfo {year} {2015})}\BibitemShut {NoStop}%
	\bibitem [{\citenamefont {Madjarov}\ \emph {et~al.}(2019)\citenamefont
		{Madjarov}, \citenamefont {Cooper}, \citenamefont {Shaw}, \citenamefont
		{Covey}, \citenamefont {Schkolnik}, \citenamefont {Yoon}, \citenamefont
		{Williams},\ and\ \citenamefont {Endres}}]{Madjarov_2019}%
	\BibitemOpen
	\bibfield  {author} {\bibinfo {author} {\bibfnamefont {I.~S.}\ \bibnamefont
			{Madjarov}}, \bibinfo {author} {\bibfnamefont {A.}~\bibnamefont {Cooper}},
		\bibinfo {author} {\bibfnamefont {A.~L.}\ \bibnamefont {Shaw}}, \bibinfo
		{author} {\bibfnamefont {J.~P.}\ \bibnamefont {Covey}}, \bibinfo {author}
		{\bibfnamefont {V.}~\bibnamefont {Schkolnik}}, \bibinfo {author}
		{\bibfnamefont {T.~H.}\ \bibnamefont {Yoon}}, \bibinfo {author}
		{\bibfnamefont {J.~R.}\ \bibnamefont {Williams}},\ and\ \bibinfo {author}
		{\bibfnamefont {M.}~\bibnamefont {Endres}},\ }\bibfield  {title} {\bibinfo
		{title} {An atomic-array optical clock with single-atom readout},\ }\href
	{https://doi.org/10.1103/PhysRevX.9.041052} {\bibfield  {journal} {\bibinfo
			{journal} {Physical Review X}\ }\textbf {\bibinfo {volume} {9}},\ \bibinfo
		{pages} {041052} (\bibinfo {year} {2019})}\BibitemShut {NoStop}%
	\bibitem [{\citenamefont {Okuno}\ \emph {et~al.}(2022)\citenamefont {Okuno},
		\citenamefont {Nakamura}, \citenamefont {Kusano}, \citenamefont {Takasu},
		\citenamefont {Takei}, \citenamefont {Konishi},\ and\ \citenamefont
		{Takahashi}}]{Okuno_2022}%
	\BibitemOpen
	\bibfield  {author} {\bibinfo {author} {\bibfnamefont {D.}~\bibnamefont
			{Okuno}}, \bibinfo {author} {\bibfnamefont {Y.}~\bibnamefont {Nakamura}},
		\bibinfo {author} {\bibfnamefont {T.}~\bibnamefont {Kusano}}, \bibinfo
		{author} {\bibfnamefont {Y.}~\bibnamefont {Takasu}}, \bibinfo {author}
		{\bibfnamefont {N.}~\bibnamefont {Takei}}, \bibinfo {author} {\bibfnamefont
			{H.}~\bibnamefont {Konishi}},\ and\ \bibinfo {author} {\bibfnamefont
			{Y.}~\bibnamefont {Takahashi}},\ }\bibfield  {title} {\bibinfo {title}
		{High-resolution spectroscopy and single-photon rydberg excitation of
			reconfigurable ytterbium atom tweezer arrays utilizing a metastable state},\
	}\bibfield  {journal} {\bibinfo  {journal} {Journal of the Physical Society
			of Japan}\ }\textbf {\bibinfo {volume} {91}},\ \href
	{https://doi.org/10.7566/JPSJ.91.084301} {10.7566/JPSJ.91.084301} (\bibinfo
	{year} {2022})\BibitemShut {NoStop}%
	\bibitem [{\citenamefont {Chiarotti}\ \emph {et~al.}(2022)\citenamefont
		{Chiarotti}, \citenamefont {Tinsley}, \citenamefont {Bandarupally},
		\citenamefont {Manzoor}, \citenamefont {Sacco}, \citenamefont {Salvi},\ and\
		\citenamefont {Poli}}]{Chiarotti_2022}%
	\BibitemOpen
	\bibfield  {author} {\bibinfo {author} {\bibfnamefont {M.}~\bibnamefont
			{Chiarotti}}, \bibinfo {author} {\bibfnamefont {J.~N.}\ \bibnamefont
			{Tinsley}}, \bibinfo {author} {\bibfnamefont {S.}~\bibnamefont
			{Bandarupally}}, \bibinfo {author} {\bibfnamefont {S.}~\bibnamefont
			{Manzoor}}, \bibinfo {author} {\bibfnamefont {M.}~\bibnamefont {Sacco}},
		\bibinfo {author} {\bibfnamefont {L.}~\bibnamefont {Salvi}},\ and\ \bibinfo
		{author} {\bibfnamefont {N.}~\bibnamefont {Poli}},\ }\bibfield  {title}
	{\bibinfo {title} {Practical limits for large-momentum-transfer clock atom
			interferometers},\ }\href {https://doi.org/10.1103/PRXQuantum.3.030348}
	{\bibfield  {journal} {\bibinfo  {journal} {PRX Quantum}\ }\textbf {\bibinfo
			{volume} {3}},\ \bibinfo {pages} {030348} (\bibinfo {year}
		{2022})}\BibitemShut {NoStop}%
	\bibitem [{\citenamefont {Matei}\ \emph {et~al.}(2017)\citenamefont {Matei},
		\citenamefont {Legero}, \citenamefont {Häfner}, \citenamefont {Grebing},
		\citenamefont {Weyrich}, \citenamefont {Zhang}, \citenamefont {Sonderhouse},
		\citenamefont {Robinson}, \citenamefont {Ye}, \citenamefont {Riehle},\ and\
		\citenamefont {Sterr}}]{Matei2017}%
	\BibitemOpen
	\bibfield  {author} {\bibinfo {author} {\bibfnamefont {D.}~\bibnamefont
			{Matei}}, \bibinfo {author} {\bibfnamefont {T.}~\bibnamefont {Legero}},
		\bibinfo {author} {\bibfnamefont {S.}~\bibnamefont {Häfner}}, \bibinfo
		{author} {\bibfnamefont {C.}~\bibnamefont {Grebing}}, \bibinfo {author}
		{\bibfnamefont {R.}~\bibnamefont {Weyrich}}, \bibinfo {author} {\bibfnamefont
			{W.}~\bibnamefont {Zhang}}, \bibinfo {author} {\bibfnamefont
			{L.}~\bibnamefont {Sonderhouse}}, \bibinfo {author} {\bibfnamefont
			{J.}~\bibnamefont {Robinson}}, \bibinfo {author} {\bibfnamefont
			{J.}~\bibnamefont {Ye}}, \bibinfo {author} {\bibfnamefont {F.}~\bibnamefont
			{Riehle}},\ and\ \bibinfo {author} {\bibfnamefont {U.}~\bibnamefont
			{Sterr}},\ }\bibfield  {title} {\bibinfo {title} {1.5 $\mu$m lasers with
			sub-10 mhz linewidth},\ }\href
	{https://doi.org/10.1103/physrevlett.118.263202} {\bibfield  {journal}
		{\bibinfo  {journal} {Physical Review Letters}\ }\textbf {\bibinfo {volume}
			{118}},\ \bibinfo {pages} {263202} (\bibinfo {year} {2017})}\BibitemShut
	{NoStop}%
	\bibitem [{\citenamefont {Oelker}\ \emph {et~al.}(2019)\citenamefont {Oelker},
		\citenamefont {Hutson}, \citenamefont {Kennedy}, \citenamefont {Sonderhouse},
		\citenamefont {Bothwell}, \citenamefont {Goban}, \citenamefont {Kedar},
		\citenamefont {Sanner}, \citenamefont {Robinson}, \citenamefont {Marti},
		\citenamefont {Matei}, \citenamefont {Legero}, \citenamefont {Giunta},
		\citenamefont {Holzwarth}, \citenamefont {Riehle}, \citenamefont {Sterr},\
		and\ \citenamefont {Ye}}]{Oelker_2019}%
	\BibitemOpen
	\bibfield  {author} {\bibinfo {author} {\bibfnamefont {E.}~\bibnamefont
			{Oelker}}, \bibinfo {author} {\bibfnamefont {R.~B.}\ \bibnamefont {Hutson}},
		\bibinfo {author} {\bibfnamefont {C.~J.}\ \bibnamefont {Kennedy}}, \bibinfo
		{author} {\bibfnamefont {L.}~\bibnamefont {Sonderhouse}}, \bibinfo {author}
		{\bibfnamefont {T.}~\bibnamefont {Bothwell}}, \bibinfo {author}
		{\bibfnamefont {A.}~\bibnamefont {Goban}}, \bibinfo {author} {\bibfnamefont
			{D.}~\bibnamefont {Kedar}}, \bibinfo {author} {\bibfnamefont
			{C.}~\bibnamefont {Sanner}}, \bibinfo {author} {\bibfnamefont {J.~M.}\
			\bibnamefont {Robinson}}, \bibinfo {author} {\bibfnamefont {G.~E.}\
			\bibnamefont {Marti}}, \bibinfo {author} {\bibfnamefont {D.~G.}\ \bibnamefont
			{Matei}}, \bibinfo {author} {\bibfnamefont {T.}~\bibnamefont {Legero}},
		\bibinfo {author} {\bibfnamefont {M.}~\bibnamefont {Giunta}}, \bibinfo
		{author} {\bibfnamefont {R.}~\bibnamefont {Holzwarth}}, \bibinfo {author}
		{\bibfnamefont {F.}~\bibnamefont {Riehle}}, \bibinfo {author} {\bibfnamefont
			{U.}~\bibnamefont {Sterr}},\ and\ \bibinfo {author} {\bibfnamefont
			{J.}~\bibnamefont {Ye}},\ }\bibfield  {title} {\bibinfo {title}
		{Demonstration of $4.8\times10^{-17}$ stability at 1 s for two independent
			optical clocks},\ }\href {https://doi.org/10.1038/s41566-019-0493-4} {\
		\textbf {\bibinfo {volume} {13}},\ \bibinfo {pages} {714} (\bibinfo {year}
		{2019})}\BibitemShut {NoStop}%
	\bibitem [{\citenamefont {Paiva}\ \emph {et~al.}(2022)\citenamefont {Paiva},
		\citenamefont {Te’eni}, \citenamefont {Peled}, \citenamefont {Cohen},\ and\
		\citenamefont {Aharonov}}]{Paiva_2022}%
	\BibitemOpen
	\bibfield  {author} {\bibinfo {author} {\bibfnamefont {I.~L.}\ \bibnamefont
			{Paiva}}, \bibinfo {author} {\bibfnamefont {A.}~\bibnamefont {Te’eni}},
		\bibinfo {author} {\bibfnamefont {B.~Y.}\ \bibnamefont {Peled}}, \bibinfo
		{author} {\bibfnamefont {E.}~\bibnamefont {Cohen}},\ and\ \bibinfo {author}
		{\bibfnamefont {Y.}~\bibnamefont {Aharonov}},\ }\bibfield  {title} {\bibinfo
		{title} {Non-inertial quantum clock frames lead to non-hermitian dynamics},\
	}\bibfield  {journal} {\bibinfo  {journal} {Communications Physics}\ }\textbf
	{\bibinfo {volume} {5}},\ \href {https://doi.org/10.1038/s42005-022-01081-0}
	{10.1038/s42005-022-01081-0} (\bibinfo {year} {2022})\BibitemShut {NoStop}%
	\bibitem [{\citenamefont {Giovannetti}\ \emph {et~al.}(2015)\citenamefont
		{Giovannetti}, \citenamefont {Lloyd},\ and\ \citenamefont
		{Maccone}}]{Giovannetti_2015}%
	\BibitemOpen
	\bibfield  {author} {\bibinfo {author} {\bibfnamefont {V.}~\bibnamefont
			{Giovannetti}}, \bibinfo {author} {\bibfnamefont {S.}~\bibnamefont {Lloyd}},\
		and\ \bibinfo {author} {\bibfnamefont {L.}~\bibnamefont {Maccone}},\
	}\bibfield  {title} {\bibinfo {title} {Quantum time},\ }\href
	{https://doi.org/10.1103/PhysRevD.92.045033} {\bibfield  {journal} {\bibinfo
			{journal} {Physical Review D}\ }\textbf {\bibinfo {volume} {92}},\ \bibinfo
		{pages} {045033} (\bibinfo {year} {2015})}\BibitemShut {NoStop}%
\end{thebibliography}

%

\end{document}